\documentclass[10pt,twocolumn,english,aps,prb,superscriptaddress,showpacs,floatfix]{revtex4-1}
\usepackage[T1]{fontenc}
\usepackage[latin9]{inputenc}
\usepackage{babel}
\usepackage{amssymb}
\usepackage{graphicx}
\usepackage{esint}

\newcommand{\vsigma}{\mbox{\boldmath $\sigma$}}

\begin{document}

\title{Topological superconducting phase in helical Shiba chains}

\author{Falko Pientka}
\affiliation{\mbox{Dahlem Center for Complex Quantum Systems and Fachbereich Physik, Freie Universit\"at Berlin, 14195 Berlin, Germany}}

\author{Leonid I. Glazman}
\affiliation{Department of Physics, Yale University, New Haven, CT 06520, USA}

\author{Felix von Oppen}
\affiliation{\mbox{Dahlem Center for Complex Quantum Systems and Fachbereich Physik, Freie Universit\"at Berlin, 14195 Berlin, Germany}}

\begin{abstract}
Recently, it has been suggested that topological superconductivity and Majorana end states can be realized in a chain of magnetic impurities on the surface of an $s$-wave superconductor when the magnetic moments form a spin helix as a result of the RKKY interaction mediated by the superconducting substrate. Here, we investigate this scenario theoretically by developing a tight-binding Bogoliubov-de Gennes description starting from the Shiba bound states induced by the individual magnetic impurities. While the resulting model Hamiltonian has similarities with the Kitaev model for one-dimensional spinless $p$-wave superconductors, there are also important differences, most notably the long-range nature of hopping and pairing as well as the complex hopping amplitudes. We use both analytical and numerical approaches to explore the consequences of these differences for the phase diagram and the localization properties of the Majorana end states when the Shiba chain is in a topological superconducting phase. 
\end{abstract}


\maketitle

\section{Introduction}

Motivated in part by possible applications to topological quantum computation \cite{kitaev03}, there is currently much interest in  condensed-matter systems which support Majorana bound states.\cite{review2,review3} Systems which are investigated in the laboratory \cite{mourik12,das12,churchill13,rokhinson12,lund,harlingen,willett,kang} are based on fractional quantum Hall systems,\cite{read00} topological insulators, \cite{fu08,fu09} semiconductor quantum wires,\cite{lutchyn10,oreg10,alicea11} or cold atoms.\cite{zoller} Starting with the seminal work of Fu and Kane,\cite{fu08,fu09} a particularly promising strategy attempts to engineer topological superconducting phases hosting Majorana bound states in hybrid structures involving conventional $s$-wave superconductors. 

Recently, it has been suggested \cite{bernevig} (see also Refs.\ \onlinecite{loss0,beenakker_magnetic,flensberg,morpurgo}) that Majorana bound states could be realized in chains of magnetic impurities placed on an $s$-wave superconductor. Magnetic impurities placed in a conventional superconductor create localized, sub-gap Shiba states.\cite{yu,shiba,rusinov,rmp} When the magnetic impurities are brought close to one another, the individual localized Shiba states hybridize and may form a band. Electrons in such a band, in turn, may hybridize with the condensate of the bulk superconductor by Andreev reflection. The properties of the band and the strength of Andreev processes depend on the magnetic structure of the impurity chain.
Assuming that the impurity spins form a helix, it is argued that the Shiba bands will effectively realize a topological superconducting phase, akin to one-dimensional spinless $p$-wave superconductors.\cite{kitaev2} Indeed, the Shiba states are effectively spin polarized and the (spin-singlet) $s$-wave Cooper pairs of the superconducting substrate can induce $p$-wave superconducting correlations in the Shiba band since neighboring impurity spins are misaligned due to the helical spin order. A particularly attractive feature of this proposal is that the presence of Majorana end states could be probed directly by scanning tunneling spectroscopy.\cite{yazdani1,yazdani2}

Following the original suggestion,\cite{bernevig} some aspects of this proposal have been investigated by a number of authors. \cite{nagaosa,loss,braunecker,franz} However, a theory making the connection to the formation and hybridization of Shiba states explicit has not yet been given. It is the purpose of the present paper to provide such a theoretical description. We show that for the case of deep Shiba states, one can derive an effective tight-binding Bogoliubov-de Gennes Hamiltonian. While this tight-binding Hamiltonian shares important features with the paradigmatic Kitaev model for one-dimensional spinless $p$-wave superconductors,\cite{kitaev2} there are also several substantial differences: (i) Both the hopping and the pairing terms are long range, having a $1/r$-power-law decay with distance as long as $r$ remains small compared to the coherence length $\xi_0$ of the host superconductor. (ii) The hopping terms generally involve complex phase factors which lead to dispersions which are asymmetric 
under momentum reversal $k \to -k$. We explore the consequences of these differences both for the phase diagram and for the localization properties of Majorana end states present when the system is in a topological phase. 

Our approach is based on the following physical picture. We start with a given static texture of the impurity spins along the chain. This texture is ultimately the result of the RKKY interaction between the impurity states as mediated by the superconducting host.\cite{bernevig,loss,braunecker,franz,foot2} It seems likely that the precise nature of the spin texture is sensitive to system-specific details such as the ratio of the impurity spacing to the Fermi wavelength of the superconductor or anisotropies of the exchange interaction at the surface of the superconductor. Hence, we consider general periodic and helical spin textures which need not be commensurate with the underlying impurity chain.
As long as the magnetic impurities are sufficiently dilute, each of them binds a pair of Shiba states with energies in the superconducting gap which are symmetric about the chemical potential pinned to the center of the gap. Overlaps between the Shiba bound states lead to hybridization and the formation of bands whose bandwidth grows with decreasing spacing between the impurities. If the impurity states are shallow, i.e., their energies are close to the superconducting gap edges, the impurity band will in general merge with the quasiparticle continuum and simply smear the gap edge. Topological superconducting states can possibly still be realized when the impurity bands at positive and negative energies become wide enough to overlap around the center of the gap, i.e., when the bandwidth becomes comparable with the superconducting gap. 

Here, we focus on the opposite limit in which the Shiba bound states are deep with energies near the center of the gap. In this case, the bands emerging from the positive- and negative-energy Shiba states start to overlap already for weak hybridization. Thus, the Shiba bands remain well separated from the quasiparticle continuum and we can derive an effective Hamiltonian within the subspace of Shiba states. As long as the bandwidth of the Shiba states remains smaller than the energy of the bare Shiba states, there are two well-separated Shiba bands and the system is in a nontopological superconducting state. Topological superconducting phases can occur when the Shiba bands overlap around the center of the gap.  Since the Shiba states are spin polarized, the induced pairing amplitude within the subspace of Shiba states is necessarily odd in momentum and hence of $p$-wave nature. If the system enters such a topological phase, there will be a $p$-wave gap at the chemical potential within the overlapping Shiba 
bands, in addition to the original $s$-wave gap of the host superconductor. However, such a $p$-wave gap does not necessarily form for arbitrary parameters despite the presence of a finite $p$-wave pairing amplitude and overlapping Shiba bands. Elucidating this nontrivial phase diagram is one of the central goals of the present paper. 

The paper is organized as follows. In Sec.\ \ref{sec:Model}, we introduce the model and discuss the formation of the spin helix. In Sec.\ \ref{sec:shiba_single}, we review the formation of Shiba states for a single magnetic impurity in an $s$-wave superconductor, employing a technique which readily generalizes to chains of magnetic impurities.
The latter are discussed in Sec.\ \ref{sec:shiba_chain}, culminating in a tight-binding Bogoliubov-de Gennes equation for deep Shiba states. The tight-binding model is employed to analyze the phase diagram in Sec.\ \ref{sec:phase} and the localization properties of the Majorana modes in Sec.\ \ref{sec:majorana}. We conclude in Sec.\ \ref{sec:conclusions} and defer some technical details to appendices.

\section{Model\label{sec:Model}}

Our starting point is the Bogoliubov-de Gennes Hamiltonian (BdG) of an $s$-wave superconductor. We assume that the superconductor is in the clean limit but hosts a chain of magnetic impurities placed at locations ${\bf r}_j$,
\begin{equation}
  {\cal H} = \xi_{\bf p}\tau_z - J\sum_j {\bf S}_i\cdot{\vsigma} \delta({\bf r}-{\bf r}_j)  + \Delta \tau_x.
  \label{Hchain}
\end{equation}
Here, ${\bf p}$ and ${\bf r}$ denote the electron's momentum and position, $\xi_{\bf p} = {\bf p}^2/2m-\mu$ with the chemical potential $\mu$, $\Delta$ is the superconducting gap, and $J$ denotes the strength of the exchange coupling between the magnetic impurity with spin $S$ and the electrons in the superconductor. The Pauli matrices $\sigma_i$ ($\tau_i$) operate in spin (particle-hole) space. The BdG Hamiltonian is written in a basis which corresponds to the four-component Nambu operator $\Psi = [\psi_\uparrow, \psi_\downarrow, \psi^\dagger_\downarrow, -\psi^\dagger_\uparrow]$ in terms of the electronic field operator $\psi_\sigma({\bf r})$. In this basis, the time-reversal operator takes the form $T = i \sigma_y K$, where $K$ denotes complex conjugation. The BdG Hamiltonian (\ref{Hsingleimp}) obeys the symmetry $\{{\cal H},CT\} = 0$, with $C = - i\tau_y$. Thus, if $\psi$ is an eigenspinor of ${\cal H}$ with energy $E$, $CT\psi$ is an eigenspinor of energy $(-E)$. 

We assume that the magnetic moments are classical and arranged along a linear chain with lattice spacing $a$. We can parametrize the impurity spins ${\bf S}_j$ through spherical coordinates, using the angles $\theta_j$ and $\phi_j$ in addition to $S$,
\begin{equation}
  {\bf S}_j = S (\sin\theta_j \cos\phi_j, \sin\theta_j \sin\phi_j, \cos\theta_j ). 
\end{equation}
In the BdG equation (\ref{Hchain}), we take the spins as frozen into a given spin texture ${\bf S}_j$. We also assume the impurity spacings $a\gg 1/k_F$ so that the bandwidth of the Shiba bands is small compared to the gap of the host superconductor.

The spin texture will in general be governed by the RKKY interaction between the impurity spins, as mediated by the superconducting host.\cite{bernevig,loss,braunecker,franz,foot2} Magnetic impurities interact with each other via exchange by virtual electron-hole excitations in the host metal. If the host is in the normal state, this exchange leads to the familiar RKKY interaction between the impurities\cite{RKKY} whose sign alternates as a function of inter-impurity distance $r_{ij}$ and which is of magnitude $J(i-j)\sim (J\nu_0)^2 v_F/(k_F^2r_{ij}^3)$. Here, $J$ denotes the exchange coupling between magnetic impurity and electrons, $\nu_0$ is the electronic density of states at the Fermi energy, and $v_F$ and $k_F$ denote the Fermi velocity and wavevector, respectively. There is some evidence for normal-metal substrates\cite{wiesendanger} that the RKKY interaction between impurity spins can lead to the formation of a spin helix when the impurities form an ordered chain.\cite{footnote}  

In a clean system, the RKKY interaction between two magnetic impurities a distance $r_{ij}$ apart involves virtual electron-hole pairs with characteristic energy $\hbar v_F/r_{ij}$. On the other hand, superconductivity prohibits pairs with energy less than the gap $\Delta$. As a result, $J(i-j)$ is substantially affected by superconductivity if $r_{ij}\gtrsim\xi_0$, where $\xi_0$ is the coherence length of the host superconductor. A perturbative treatment of the exchange interaction indicates\cite{RKKY,galitski} that the correction to the normal-state value of $J(i-j)$ caused by superconductivity is antiferromagnetic. The magnitude of the correction is of order $\delta J(i-j)\sim \Delta/(k_Fr_{ij})^2$ and thus small at any  $r_{ij}\lesssim\xi_0$.
The presence of deep Shiba states enhances the correction;\cite{yao} in the limit $\alpha\to 1$, its absolute value reaches a maximum of the order of $\Delta/(k_Fr_{ij})$. [This estimate may be obtained from a consideration of the total energy of a superconductor containing a pair of impurities which create Shiba states according to Eqs.~(23), (32), and (33) below.] Still, the ``normal-state'' RKKY wins over the correction at $k_Fr_{ij}\lesssim (k_Fv_F/\Delta)^{1/2}$. For a typical superconductor, the right-hand side gives a relatively mild limitation $\sim 10^2$ which is compatible with the assumption $k_Fa\gg 1$. Note that at the border of that region, $k_Fr_{ij}\sim (k_Fv_F/\Delta)^{1/2}$, the energy scale important for the observation of Majorana states, $\Delta_{\rm eff}$, is already small, $\Delta_{\rm eff}\sim\Delta^{3/2}/(k_Fv_F)^{1/2}$. Placing the magnetic impurities closer to each other makes both, the band of Shiba states and the induced gap wider. At the same time, superconductivity of the host 
will hardly affect the mutual orientation of the magnetic moments.

In general, one would expect that the specifics of the spin helix such as the overall spin orientation or the pitch depend sensitively on the details of the system. Important parameters are the ratio of the impurity spacing and the Fermi wavelength of the host superconductor, the single-ion magnetic anisotropy, as well as the spatial structure and isotropy of the exchange interaction between the magnetic moments. For this reason, we consider a general class of helical spin textures of the form
\begin{equation}
  \theta_j = \theta \,\,\,\,\,  ; \,\,\,\,\, \phi_j = 2 k_h x_j
\label{spinhelix} 
\end{equation} 
with a constant opening angle $\theta$ and pitch $\pi/k_h$; here $x_j=ja$ denotes the position of the $j$-th impurity along the chain. 

The possible values of $k_h$ are determined by the maxima of the Fourier transform of $J(i-j)$.\cite{foot2,loss,braunecker,franz} Thus, for a simple isotropic model of the superconductor and at $k_Fa\gg 1$, the RKKY interaction $J(i-j)$ results in a helix wavevector $2k_h=(2k_Fa-2\pi n)/a$ with a single value of $n$ such that $|k_ha|\leq\pi$. While we investigate this simple case in some detail below, we will first discuss the phase diagram and the Majorana bound states for arbitrary spin helices as defined in Eq.\ (\ref{spinhelix}). The reason is that the details of the band structure of the superconductor as well as possible spin-orbit coupling may allow for other relations between the Fermi wavevector $k_F$ and the helix wavevector $k_h$. Note also that if the Hamiltonian of the magnetic system is dominated by the exchange interaction with an isotropic exchange integral $J(i-j)$,\cite{galitski} the spin helix is planar with $\theta=\pi/2$.
In general, the value of the single-ion anisotropy depends strongly on the orbital moment of the magnetic ion and the coordination of the host lattice.\cite{anisotropy,fert,kern} 

\section{Shiba states\label{sec:shiba}}

\subsection{Single magnetic impurity}
\label{sec:shiba_single}

To provide necessary background and to fix notation, we briefly derive the Shiba states for a single magnetic impurity in a form which can be generalized to a chain of impurities. For a single impurity placed at the origin, the BdG Hamiltonian (\ref{Hchain}) simplifies to
\begin{equation}
  {\cal H} = \xi_{\bf p}\tau_z - J{\bf S}\cdot{\vsigma} \delta({\bf r})  + \Delta \tau_x.
  \label{Hsingleimp}
\end{equation}
We can choose the impurity spin ${\bf S}$ to point along the $z$ direction. In this case, the $4\times 4$ Hamiltonian in Eq.\ (\ref{Hsingleimp}) separates into independent $2\times2$ Hamiltonians ${\cal H}_\pm$ for spin-up ($+$) and spin-down ($-$) electrons,
\begin{equation}
 {\cal H}_\pm = \xi_{\bf p} \tau_z \mp J S \delta({\bf r}) + \Delta \tau_x.
 \label{Hpm}
\end{equation}
To solve for the bound-state spectrum of these Hamiltonians, we write the BdG equations in a way which isolates the impurity term on the right-hand side,
\begin{equation}
    [E - \xi_{\bf p} \tau_z - \Delta \tau_x] \psi({\bf r}) = \mp J S \delta({\bf r}) \psi({\bf 0}),
\end{equation}  
and pass to momentum space using $\psi({\bf r}) = \int [d{\bf p}/(2\pi)^3] e^{i{\bf p}\cdot{\bf r}}\psi_{\bf p}$, 
\begin{equation}
    [E - \xi_{\bf p} \tau_z - \Delta \tau_x] \psi_{\bf p} = \mp J S \psi({\bf 0}).
\end{equation}  
Multiplying by $[E - \xi_{\bf p} \tau_z - \Delta \tau_x]^{-1}$ from the left, we obtain
\begin{equation}
   \psi_{\bf p} = \frac{\mp J S}{E^2 - \xi_{\bf p}^2 -\Delta^2} [E + \xi_{\bf p} \tau_z + \Delta \tau_x] \psi({\bf 0}).
\end{equation}
We can now turn this into an equation for the spinor $\psi({\bf 0})$ evaluated at the position of the impurity only,
\begin{equation}
   \psi({\bf 0}) =  \int \frac{d{\bf p}}{(2\pi)^3}\frac{\mp J S}{E^2 - \xi_{\bf p}^2 -\Delta^2} [E + \xi_{\bf p} \tau_z + \Delta \tau_x] \psi({\bf 0}).
\end{equation}
The integral is readily evaluated (see Appendix \ref{integrals}) and we obtain a linear set of equations for the BdG spinor at the position of the impurity,
\begin{equation}
   \left\{ {\bf 1} \mp \frac{\alpha}{\sqrt{\Delta^2 - E^2}} [E + \Delta \tau_x]\right\} \psi({\bf 0}) = 0.
  \label{linear}
\end{equation}
Here we introduced the dimensionless impurity strength $\alpha = \pi \nu_0 J S$ in terms of the normal-phase density of states $\nu_0$. 

One readily finds from Eq.\ (\ref{linear}) that $H_\pm$ has a subgap solution at energy $\pm E_0$ with\cite{yu,shiba,rusinov,rmp}
\begin{equation}
  E_0 = \Delta \frac{1-\alpha^2}{1+\alpha^2}.
\end{equation}
The energies of the two Shiba states cross at $\alpha = 1$ where the ground state changes from even to odd electron number. 

The corresponding eigenspinors (written in the four-spinor form of the original $4\times4$ BdG Hamiltonian)
\begin{equation}
  \psi_+({\bf 0}) = \frac{1}{\sqrt{{\cal N}}} \left( \begin{array}{c} 1 \cr 0 \cr 1 \cr 0 \end{array}\right)  \,\,\,\,\, ; \,\,\,\,\,\,\,  \psi_-({\bf 0}) = \frac{1}{\sqrt{{\cal N}}} \left(\begin{array}{c} 0 \\ 1 \\ 0 \\-1 \end{array}\right).  
  \label{zspinor}
\end{equation}
Here, the normalization factor ${\cal N} = (1+\alpha^2)^2 / 2\pi \nu_0 \Delta \alpha$ follows from the normalization condition $1 = \int d{\bf r} \sum_n |\psi_n({\bf r})|^2 = \int [d{\bf p}/(2\pi)^3] \sum_n |(\psi_{\bf p})_n|^2$. Note that the solution starting out at positive energies for small exchange coupling corresponds to quasiparticles made up from spin-up electrons and spin-down holes, while the one which starts at negative energies consists of spin-down electrons and spin-up holes. 

For later reference, it is useful to generalize these spinors to impurity spins pointing in arbitrary directions. Parameterizing the impurity spin in spherical coordinates, ${\bf S} = S (\sin\theta \cos\phi,\sin\theta \sin\phi,\cos\theta)$, the corresponding spin-up and spin-down Pauli spinors are 
\begin{equation}
  |\uparrow\rangle = \left( \begin{array}{c} \cos(\theta/2) \cr \sin(\theta/2) e^{i\phi}  \end{array}\right)   \,\,\, ; \,\,\,\,
   |\downarrow\rangle = \left( \begin{array}{c} \sin(\theta/2) e^{-i\phi} \cr -\cos(\theta/2)   \end{array}\right).
\end{equation}
In terms of these Pauli spinors, the BdG spinors in Eq.\ (\ref{zspinor}) generalize to
\begin{equation}
  \psi_+({\bf 0}) = \frac{1}{\sqrt{{\cal N}}} \left( \begin{array}{c} |\uparrow\rangle \cr |\uparrow\rangle \end{array}\right)  \,\,\,\,\, ; \,\,\,\,\,\,\,  \psi_-({\bf 0}) = \frac{1}{\sqrt{{\cal N}}} \left(\begin{array}{c} |\downarrow\rangle \\ -|\downarrow\rangle \end{array}\right).  
\label{spinor}
\end{equation}
Note that the Pauli spinors are related by time reversal symmetry, $|\downarrow\rangle = T |\uparrow\rangle$, so that the BdG spinors satisfy the relation $\psi_-({\bf 0}) = CT \psi_+({\bf 0})$ in accordance with the general symmetries of the BdG Hamiltonian. 

\subsection{Chain of magnetic impurities}
\label{sec:shiba_chain}

\subsubsection{General formulation}

We now generalize the approach of the previous section to a chain of magnetic impurities ${\bf S}_j$ at sites ${\bf r}_j$ as described by the Hamiltonian in Eq.\ (\ref{Hchain}). As for a single impurity, we start by isolating the impurity terms on one side of the BdG equation and passing to momentum space. This yields
\begin{equation}
    [E - \xi_{\bf p} \tau_z - \Delta \tau_x] \psi_{\bf p} = - J \sum_j   {\bf S}_j\cdot{\vsigma} e^{-i{\bf p}\cdot {\bf r}_j} \psi({\bf r}_j).
\end{equation}  
Multiplying from the left by $[E - \xi_{\bf p} \tau_z - \Delta \tau_x]^{-1}$ and evaluating $\psi({\bf r}_i)$ yields a closed set of equations for the BdG spinors at the positions of the impurities, 
\begin{equation}
   \psi({\bf r}_i) = - J \sum_j \int \frac{d{\bf p}}{(2\pi)^3} \frac{e^{i{\bf p}({\bf r}_i-{\bf r}_j)}}{E - \xi_{\bf p} \tau_z - \Delta \tau_x} {\bf S}_j\cdot{\vsigma} \psi({\bf r}_j).
\end{equation}
We are searching for subgap states so that we need to evaluate the momentum integral on the RHS for energies $E<\Delta$. This integral is performed in Appendix \ref{integrals} and we find
\begin{equation}
   \psi({\bf r}_i) = -  \sum_j  J_E({\bf r}_i - {\bf r}_j) {\bf\hat S}_j\cdot{\vsigma} \psi({\bf r}_j),   
 \label{LinearChain}
\end{equation}
where we defined the unit vector ${\bf\hat S}_j = {\bf S}_j/S$ and
\begin{widetext}
\begin{equation}
 J_E({\bf r}) = - \frac{\alpha}{\sqrt{\Delta^2 - E^2}} \frac{e^{-r/\xi_E}}{k_F r} \left(\begin{array}{cc} E\sin k_F r + \sqrt{\Delta^2 -E^2}\cos k_F    r & \Delta \sin k_Fr \cr \Delta \sin k_F r & E\sin k_F r - \sqrt{\Delta^2 -E^2} \cos k_F r \end{array} \right)
\label{coupling}
\end{equation}
\end{widetext}
in terms of $\xi_E = v_F/\sqrt{\Delta^2 - E^2}$. 

\subsubsection{Tight-binding model for deep impurities}

We now specify to deep impurities with impurity strength $\alpha$ close to unity so that the energy $\epsilon_0 \simeq \Delta(1-\alpha)$ of the individual Shiba states is close to the center of the gap. Moreover, we assume that the impurities are sufficiently dilute that the resulting impurity band remains well within the superconducting gap. In this limit, we can expand to linear order in $E$ [and hence in $(1-\alpha)$] as well as in the coupling between impurity sites. 

We start by writing Eq.\ (\ref{LinearChain}) as
\begin{equation}
   \psi({\bf r}_i) + J_E({\bf 0}) {\bf\hat S}_i\cdot{\vsigma} \psi({\bf r}_i)= -  \sum_{j\neq i}  J_E({\bf r}_{ij}) {\bf\hat S}_j\cdot{\vsigma} \psi({\bf r}_j),  
\end{equation}
with the shorthand ${\bf r}_{ij} = {\bf r}_i - {\bf r}_j$. The RHS is already linear in the coupling between Shiba states so that we can evaluate it for $E=0$ and $\alpha=1$. The LHS is readily expanded using Eq.\ (\ref{coupling}), so that we obtain
\begin{widetext}
\begin{equation}
\{ {\bf 1} - [E/\Delta + \alpha \tau_x] {\bf\hat S}_i\cdot{\vsigma}\} \psi({\bf r}_i) 
 =  \sum_{j\neq i} \frac{e^{-r_{ij}/\xi_0}}{k_F r_{ij}} 
  [\tau_z \cos k_F r_{ij} + \tau_x \sin k_F r_{ij} ] {\bf\hat S}_j\cdot{\vsigma} \psi({\bf r}_j),   
\end{equation}
Multiplying by ${\bf\hat S}_i\cdot{\vsigma}$ and using the identity $({\bf\hat S}_i\cdot{\vsigma})({\bf\hat S}_i\cdot{\vsigma}) = 1$ yields
\begin{equation}
\{ {\bf\hat S}_i\cdot{\vsigma} - [E/\Delta + \alpha \tau_x]\}  \psi({\bf r}_i) 
 =   \sum_{j\neq i} \frac{e^{-r_{ij}/\xi_0}}{k_F r_{ij}} 
  [\tau_z \cos k_F r_{ij} + \tau_x \sin k_F r_{ij} ] ({\bf\hat S}_i\cdot{\vsigma})({\bf\hat S}_j\cdot{\vsigma}) \psi({\bf r}_j),
  \label{effequation}   
\end{equation}
We can now project this equation to the set of Shiba states in Eq.\ (\ref{spinor}) localized at the impurities. If there are $N$ impurities, the resulting equation is a tight-binding model with a $2N\times 2N$ Hamiltonian which takes the form of a BdG equation,
\begin{equation}
 \tilde   H_{\rm eff} \phi = E \phi, 
\end{equation}
with an effective Hamiltonian
\begin{equation}
 \tilde H_{\rm eff} = \left( \begin{array} {cc}  \tilde h_{\rm eff} & \tilde \Delta_{\rm eff} \cr  \tilde \Delta_{\rm eff}^\dagger & - \tilde h_{\rm eff}^T  \end{array} \right).
\end{equation}
Here,  $\tilde h_{\rm eff}^T$ denotes the time reverse of $\tilde h_{\rm eff}$. Taking matrix elements of Eq.\ (\ref{effequation}), the entries of the effective Hamiltonian $\tilde H_{\rm eff}$ take the form
\begin{equation}
  (\tilde h_{\rm eff})_{ij} = \left\{ \begin{array}{lcc} \epsilon_0  & \,\,\,\,\,\, & i=j \cr 
     -\Delta \frac{\sin k_F r_{ij}}{k_F r_{ij}} e^{-r_{ij}/\xi_0} \langle \uparrow, i| \uparrow, j\rangle & & i\neq j . \end{array}
     \right.
\end{equation}
and 
\begin{equation}
  (\tilde \Delta_{\rm eff})_{ij} = \left\{ \begin{array}{lcc} 0  & \,\,\,\,\,\, & i=j \cr 
     \Delta \frac{\cos k_F r_{ij}}{k_F r_{ij}} e^{-r_{ij}/\xi_0} \langle \uparrow, i| \downarrow, j\rangle & & i\neq j . \end{array}
     \right.
\end{equation}
In these expressions, the electronic spin states $|\sigma,i\rangle$ correspond to spin $\sigma=\uparrow,\downarrow$ with respect to the direction of the $i$th impurity spin. Parameterizing these impurity spin directions through angles $\theta_i$ and $\phi_i$, we have
\begin{eqnarray}
   \langle \uparrow, i| \uparrow, j\rangle &=& \cos \frac{\theta_i}{2} \cos \frac{\theta_j}{2} + \sin \frac{\theta_i}{2}\sin \frac{\theta_j}{2}
    e^{i(\phi_j-\phi_i)}
   \\
   \langle \uparrow, i| \downarrow, j\rangle &=& e^{-i(\phi_i+\phi_j)/2}\left[\cos \frac{\theta_i}{2} \sin \frac{\theta_j}{2} e^{-i(\phi_j-\phi_i)/2} - \sin    \frac{\theta_i}{2}\cos \frac{\theta_j}{2} e^{-i(\phi_i-\phi_j)/2}\right]
\end{eqnarray}
Note that the pairing terms involve a site-dependent phase factor $\exp\{- i(\phi_i+\phi_j)/2\}$. It is convenient to eliminate this phase factor by a gauge transformation 
\begin{equation}
 {\cal U} = \left( \begin{array}{cc} e^{i\phi/2} & 0 \cr 0 & e^{-i\phi/2} \end{array}\right) ,
\end{equation}
where $\phi$ denotes a matrix in site space with matrix elements $\phi_{ij} = \delta_{ij} \phi_j$. Performing this unitary transformation, we find the effective Hamiltonian    
\begin{equation}
     {\cal H} = {\cal U}\tilde{\cal H}{\cal U}^\dagger = \left( \begin{array} {cc}  h_{\rm eff} &  \Delta_{\rm eff} \cr   \Delta_{\rm eff}^\dagger & -       h_{\rm eff}^T  \end{array} \right)
\end{equation}
with
\begin{equation}
  (h_{\rm eff})_{ij} = \left\{ \begin{array}{lcc} \epsilon_0  & \,\,\,\,\,\, & i=j \cr 
     -\Delta \frac{\sin k_F r_{ij}}{k_F r_{ij}} e^{-r_{ij}/\xi_0} \left[\cos \frac{\theta_i}{2} \cos \frac{\theta_j}{2} e^{i(\phi_i-\phi_j)/2} + \sin \frac{\theta_i}{2}\sin \frac{\theta_j}{2}
    e^{-i(\phi_i-\phi_j)/2}\right]  & & i\neq j . \end{array}
     \right.
\end{equation}
and 
\begin{equation}
  (\Delta_{\rm eff})_{ij} = \left\{ \begin{array}{lcc} 0  & \,\,\,\,\,\, & i=j \cr 
     \Delta \frac{\cos k_F r_{ij}}{k_F r_{ij}} e^{-r_{ij}/\xi_0} \left[ \cos \frac{\theta_i}{2} \sin \frac{\theta_j}{2} e^{i(\phi_i-\phi_j)/2} - \sin    \frac{\theta_i}{2}\cos \frac{\theta_j}{2} e^{-i(\phi_i-\phi_j)/2}   \right] & & i\neq j . \end{array}
     \right. .
\end{equation}
Finally, we specify the Hamiltonian to a spin helix as defined in Eq.\ (\ref{spinhelix}) and find 
\begin{equation}
  (h_{\rm eff})_{ij} = \left\{ \begin{array}{lcc} \epsilon_0  & \,\,\,\,\,\, & i=j \cr 
     -\Delta \frac{\sin k_F r_{ij}}{k_F r_{ij}} e^{-r_{ij}/\xi_0} \left[  e^{ik_hx_{ij}} \cos^2 \frac{\theta}{2} +  
     e^{-ik_hx_{ij}} \sin^2  \frac{\theta}{2} \right]  & & i\neq j  \end{array}
     \right.
     \label{heff}
\end{equation}
as well as
\begin{equation}
  (\Delta_{\rm eff})_{ij} = \left\{ \begin{array}{lcc} 0  & \,\,\,\,\,\, & i=j \cr 
     i \Delta \frac{\cos k_F r_{ij}}{k_F r_{ij}} e^{-r_{ij}/\xi_0}  \sin \theta  \sin{k_hx_{ij}}  & & i\neq j , \end{array}
     \right. 
     \label{deltaeff}
\end{equation}
\end{widetext}
where we use the notation $x_{ij}= x_i - x_j$. Note that in this form, the Hamiltonian is translationally invariant. Conveniently, $\epsilon_0$ only enters the onsite terms and $-\epsilon_0$ effectively acts as a chemical potential for the band of Shiba states. While this BdG Hamiltonian is reminiscent of the Kitaev chain,\cite{kitaev2} there are several characteristic differences:
\begin{itemize}
\item The Hamiltonian involves long-range hopping terms. This is a consequence of the fact that the wavefunctions of the Shiba states fall off as $1/r$ with distance from the magnetic impurity as long as $r$ is small compared to the superconducting coherence length. 

\item In general, the hopping terms involve complex phase factors. This reflects that there are (spin) supercurrents flowing in response to the spatially varying Zeeman field of the magnetic impurities, similar to the magneto-Josephson effect. \cite{jiang,pientka} These supercurrents induce a spatially varying phase of the effective $p$-wave paring strength which we then eliminated by the unitary transformation at the expense of introducing complex phase factors into the hopping terms.\cite{romito}

\item The hopping amplitudes are real for a strictly planar spin helix with $\theta =\pi/2$ due to the additional reflection symmetry present in this case. This simplifies the site-off-diagonal hopping and pairing terms of the tight-binding Hamiltonian in Eqs.\ (\ref{heff}) and (\ref{deltaeff}), which become 
\begin{equation}
  (h_{\rm eff})_{ij} =  
     - \Delta\frac{\sin k_F r_{ij} }{k_F r_{ij}}e^{-r_{ij}/\xi_0}  \cos{k_hx_{ij}}  
     \label{heff_pi}
\end{equation}
as well as
\begin{equation}
  (\Delta_{\rm eff})_{ij} =  
      i \Delta \frac{ \cos k_F r_{ij}}{k_F r_{ij}} e^{-r_{ij}/\xi_0}  \sin{k_h x_{ij}}  
     \label{deltaeff__pi}
\end{equation}

\item While the pairing is odd and hence $p$-wave, it also involves long-range contributions which fall off as $1/r$ as long as $r$ is small compared to the superconducting coherence length.  
\end{itemize}
In the following, we explore the consequences of these differences for the phase diagram and the splitting of Majorana end states using both analytical and numerical approaches. 

\section{Phase diagram}
\label{sec:phase}

We first consider an infinite chain of Shiba states and qualitatively explore the phase diagram of the effective tight-binding model. For an infinite chain, the Hamiltonian defined by Eqs.\ (\ref{heff}) and (\ref{deltaeff}) is translationally invariant and can be solved by passing to momentum states. This yields the $2\times2$ BdG Hamiltonian
\begin{equation}
     {\cal H} = \left( \begin{array} {cc}  h_k &  \Delta_k \cr   \Delta_k^* & -       h_{-k}^*  \end{array} \right).
\label{BdGk}
\end{equation}
Here, we introduced the Fourier transforms 
\begin{equation}
 h_k = \sum_j (h_{\rm eff})_{ij}e^{ikx_{ij}}
 \label{hfourier}
\end{equation}
and 
\begin{equation}
 \Delta_k = \sum_j (\Delta_{\rm eff})_{ij}e^{ikx_{ij}}. 
\label{deltafourier}
\end{equation}
As detailed in Appendix \ref{hdelta}, the Fourier transforms can be performed explicitly and we find
\begin{equation}
   h_k = \epsilon_0 + \frac{\Delta}{k_F a}\biggl[F(k+k_h) \cos^2\frac{\theta}{2} + F(k-k_h) \sin^2\frac{\theta}{2}\biggr]
   \label{hk}
\end{equation}  
in terms of the function
\begin{eqnarray}
  F(k)  &=& - \left[ \arctan \frac{e^{-a/\xi_0}\sin(k_F +k)a}{1-e^{a/\xi_0}\cos(k_F +k)a} \right.
  \nonumber\\
  && \,\,\,\, + \left.
  \arctan\frac{e^{-a/\xi_0}\sin(k_F - k)a}{1-e^{a/\xi_0}\cos(k_F - k)a} \right] 
\label{Fk}
\end{eqnarray}
as well as 
\begin{eqnarray}
 \Delta_k &=&  \frac{\Delta \sin\theta}{4 k_F a} \left[ f(k_F + k_h + k) - f(k_F + k_h - k)   \right.
 \nonumber\\
 && \left.  \,\,\,\, - f(k_F - k_h + k) + f(k_F - k_h - k)   \right]
\label{deltak}
\end{eqnarray}
in terms of 
\begin{equation}
  f(k) = -\ln \left[ 1 + e^{-2a/\xi_0} - 2 e^{-a/\xi_0} \cos ka\right] .
  \label{fk}
\end{equation}
Both $h_k$ and $\Delta_k$ depend sensitively on the superconducting coherence length. For $\xi_0$ small or of order $a$, the hopping and pairing amplitudes are essentially local. In contrast, the slow power-law decay with $r_{ij}$ becomes relevant for large $\xi_0 \gg a$. In the following, we discuss these two limits separately.  

\subsection{Small coherence length}

We first specify the problem to the limit of small coherence lengths $\xi_0/a \ll 1$. While this limit is presumably not very relevant experimentally, it is helpful in understanding the more realistic case $\xi_0/a \gg 1$ discussed in Sec.\ \ref{large}. For simplicity, we also assume $k_h a \ll 1$ which allows us to expand both $h_k$ and $\Delta_k$ to linear order in this parameter.\cite{foot1} This yields
\begin{eqnarray}
  h_k &\simeq& \epsilon_0 - \frac{2\Delta}{k_F a}e^{-a/\xi_0} \sin k_F a
  \nonumber\\
  && \,\,\,\,\,\,\,\,\,\,\,\, \times \left[ \cos ka - {k_h a} \cos\theta \sin ka \right]
\end{eqnarray}
as well as 
\begin{equation}
   \Delta_k = \frac{2\Delta}{k_F a} e^{-a/\xi_0} (k_h a) \sin\theta \cos k_F a \sin ka . 
\end{equation}
There are several noteworthy features of $h_k$ and $\Delta_k$: The scale of the effective bandwidth of the band of Shiba states is set by $t = (\Delta/k_F a) e^{-a/\xi_0}$. By comparison, the corresponding scale $\delta = 2t (k_h a)$ for the pairing strength is parametrically smaller by a factor of $k_h a$. It is important to note that $h_k$ is asymmetric under $k \to -k$ unless $\theta = \pi/2$ and that the magnitude of the antisymmetric term is of the same order as the pairing. This asymmetry breaks the resonance between $k$ and $-k$ states and hence suppresses Cooper pairing.

\begin{figure}[t]
\begin{centering}
\includegraphics[width=.23\textwidth]{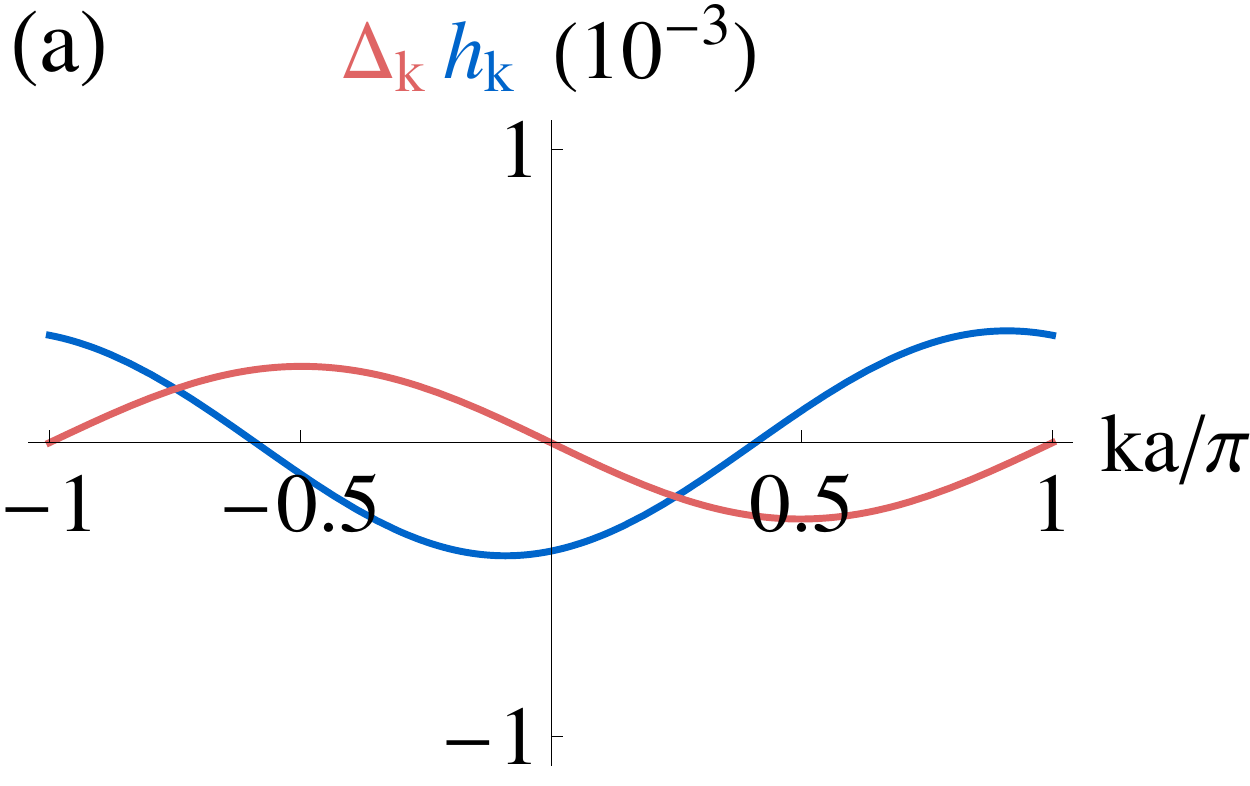}
\includegraphics[width=.23\textwidth]{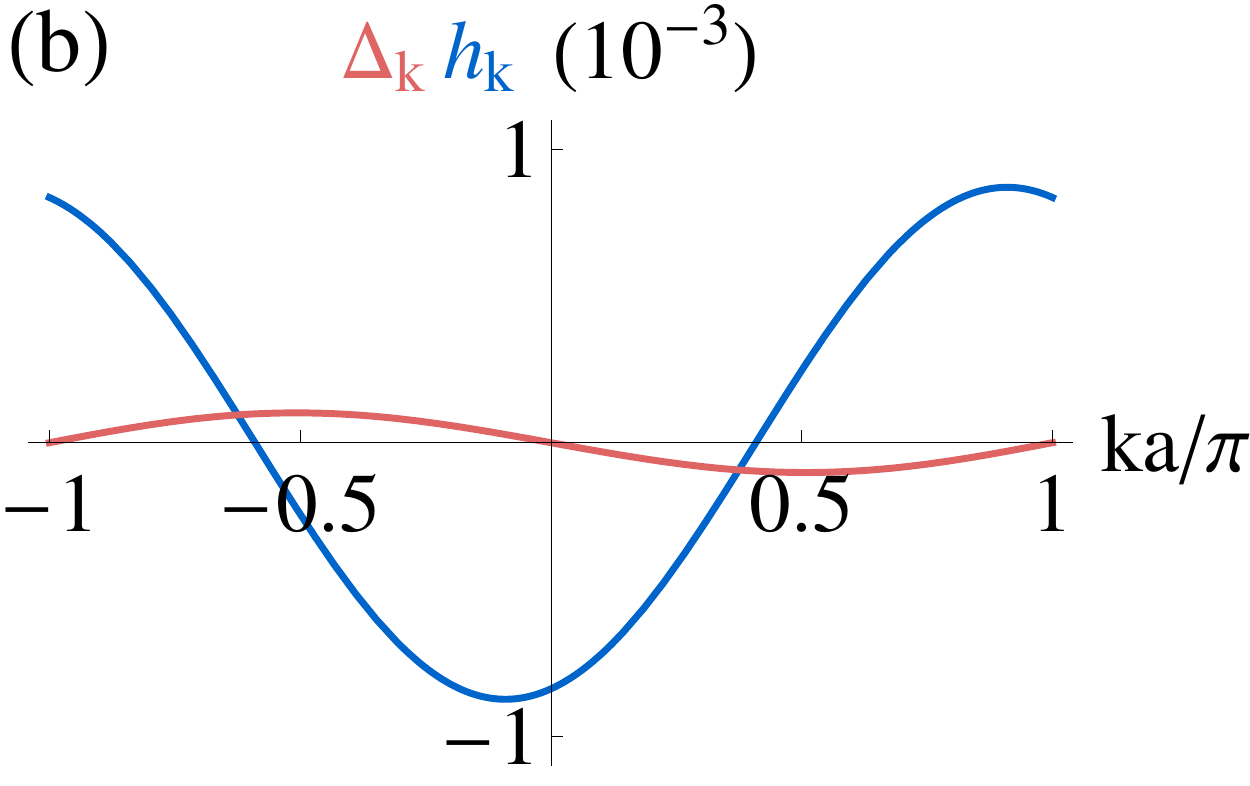}\\
\includegraphics[width=.23\textwidth]{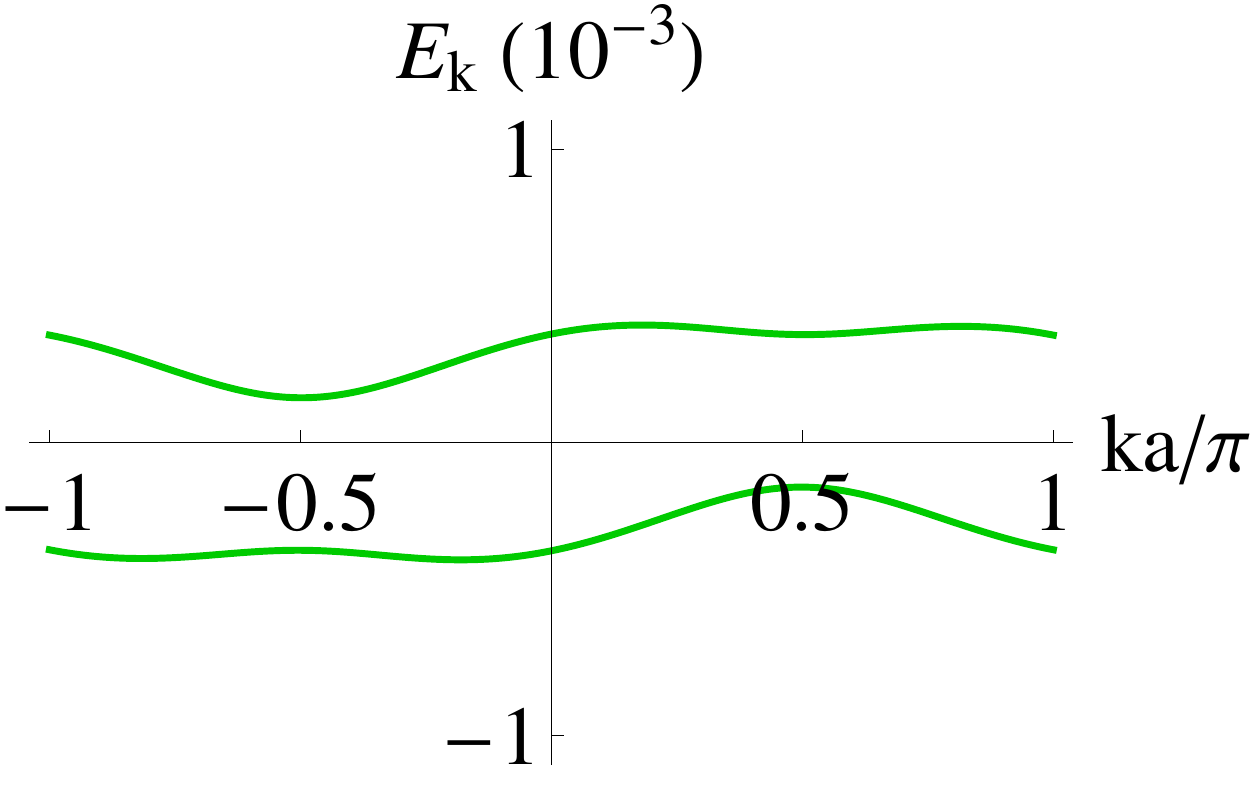}
\includegraphics[width=.23\textwidth]{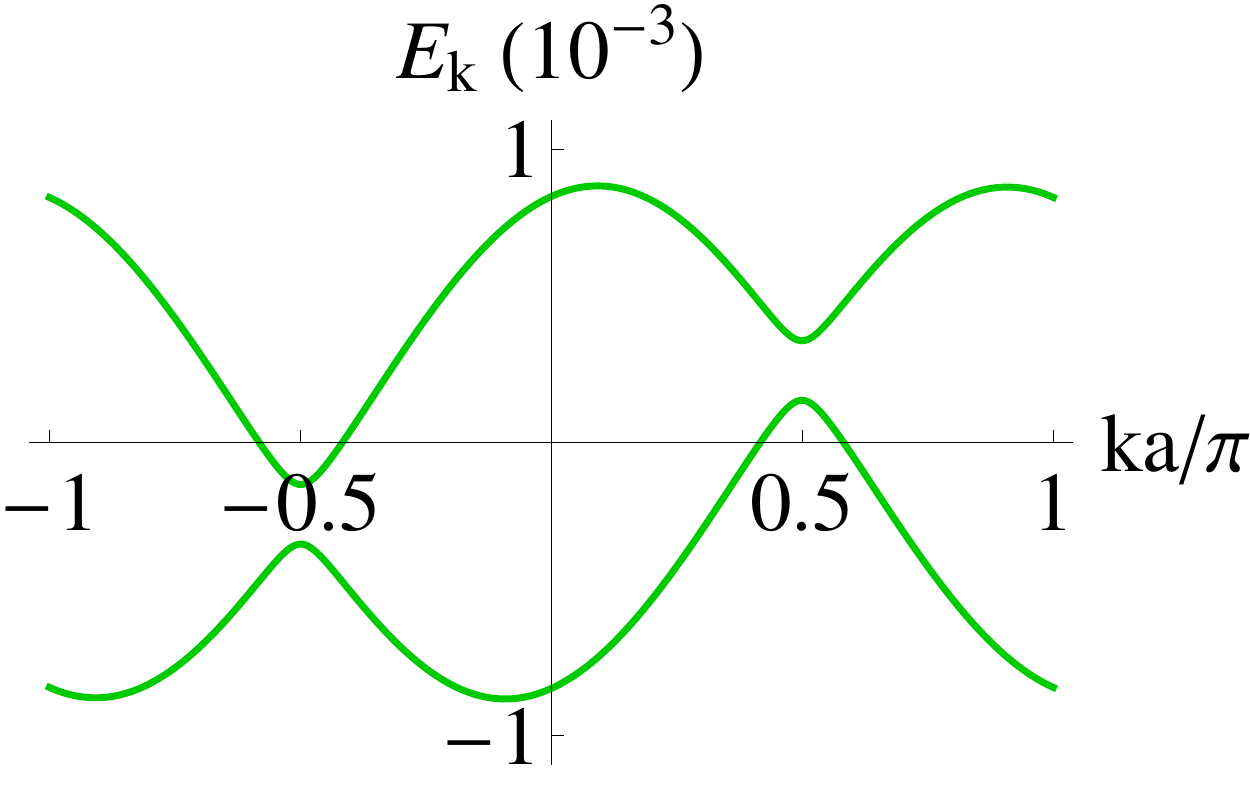}
\par\end{centering}
\caption{Numerical results for $h_k$ and $\Delta_k$ vs.\ momentum $k$ (upper panels) and the corresponding quasiparticle excitation spectra $E_k$ (lower panels) for a short coherence length $\xi_0/a=0.2$, $\epsilon_0=0$, $\theta=\pi/4$, $k_ha=\pi/8$. The plots are for $k_F a=4\pi+\pi/8$ in (a) and $k_F a=4\pi+3\pi/8$ in (b), illustrating the transitions between topological and gapless phases as a function of $k_F$. All energies are measured in units of $\Delta$.}\label{fig:small_xi}
\end{figure} 

\begin{figure}[t]
\begin{centering}
\includegraphics[width=.4\textwidth]{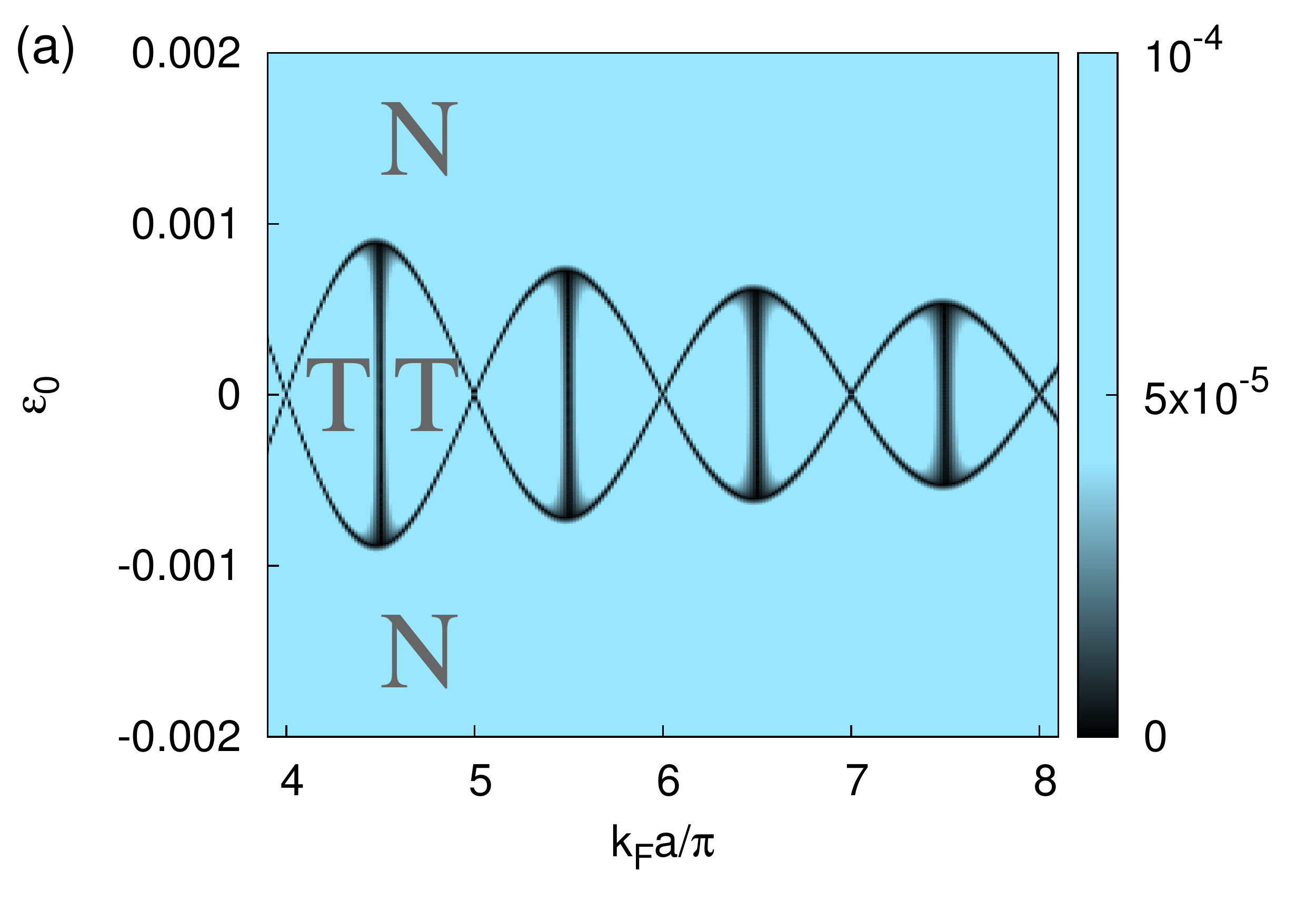} \\
\includegraphics[width=.4\textwidth]{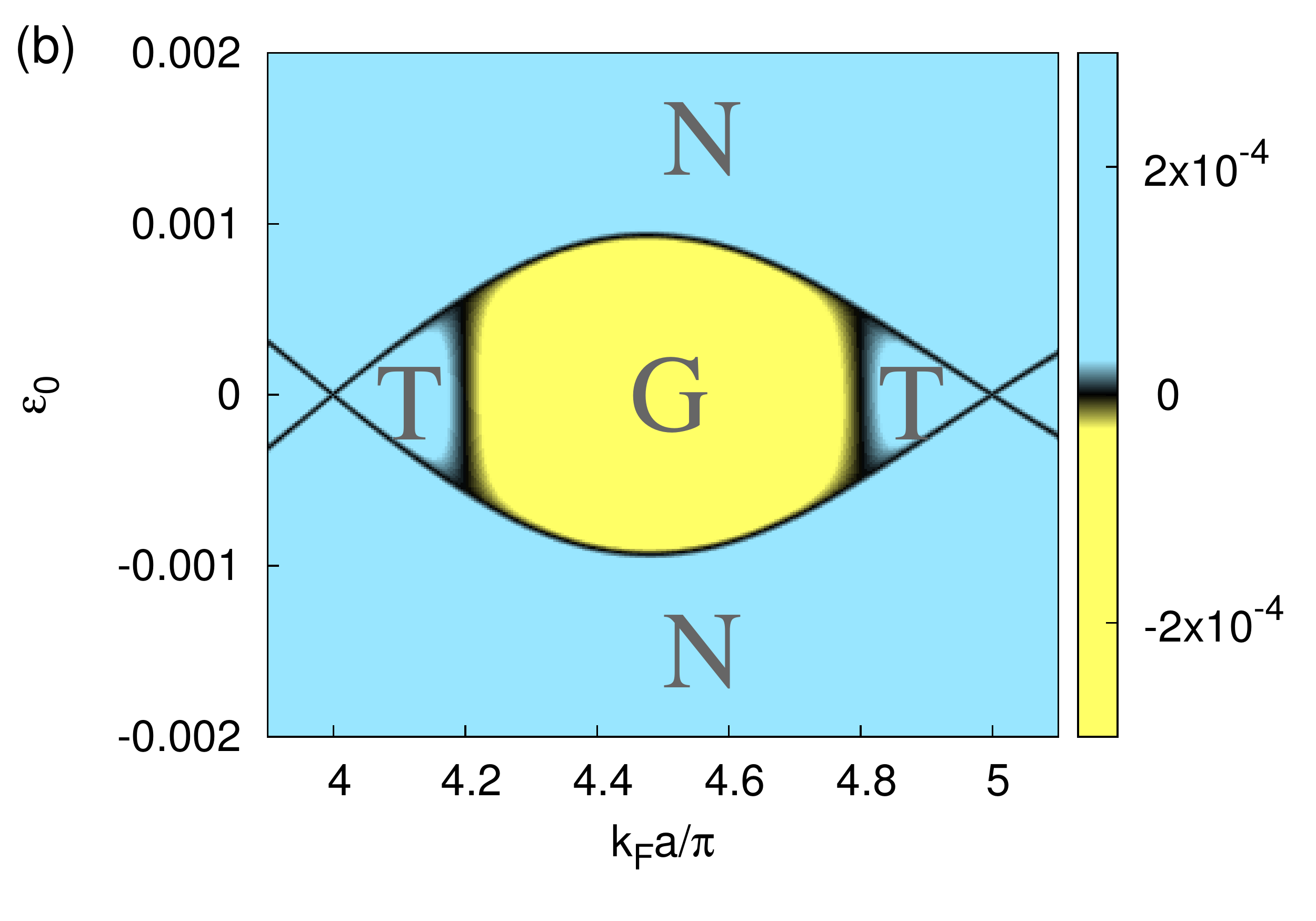}
\par\end{centering}
\caption{Numerical results for the energy minimum of the upper band (color scale) vs.\ $k_Fa$ and $\epsilon_0$ for a short coherence length $\xi_0=a/5$, $k_ha=\pi/8$, $\Delta=1$, and (a) $\theta=\pi/2$, (b) $\theta=\pi/5$. The color scale has been chosen to highlight zeros of the band minumum (black regions), which indicate topological phase transitions. The light blue regions correspond to gapped phases, while yellow regions mark the gapless phase (G). We have identified the topological (T) and nontopological (N) gapped phases using the arguments in the main text as well as by checking that a single Majorana bound state exists at both ends of the wire. In (a) the band is symmetric under $k\to -k$ and the band minimum is always nonnegative. The topological phase is centered around $\epsilon_0=0$ and the transition to the nontopological phase is approximately described by $\epsilon_0=\pm 2\sin k_Fa\, e^{-a/\xi_0}\cos k_ha/k_Fa$. The topological phase is split in half by a vertical metallic line ($\Delta_k=0$) 
at $k_Fa\simeq n\pi+\pi/2$. At $k_Fa=n\pi$ all hopping terms vanish and there can be no topological phase. In (b), the asymmetry of the spectrum expands the metallic line into a gapless phase.
}\label{fig:small_xi_PD}
\end{figure}

Diagonalizing the BdG equation (\ref{BdGk}), we thus find the subgap spectrum,
\begin{eqnarray}
  &&E_{k,\pm} = \delta \sin k_Fa \cos\theta \sin ka 
    \\
    &&\,\,\,\pm \sqrt{(\epsilon_0 - 2t \sin k_F a\cos ka)^2 + (\delta \cos k_F a \sin\theta\sin ka)^2}.
    \nonumber
\end{eqnarray}
As the energy $\epsilon_0$ of the Shiba states is reduced, the Shiba bands start to overlap and undergo a phase transition into a topological superconducting phase for appropriate parameters. Specifically, the Shiba bands cross the chemical potential at $\pm k_0$ determined by $\epsilon_0 = 2t \sin k_F a\cos k_0a$. The pairing term opens $p$-wave gaps at $\pm k_0$ of magnitude $\delta |\cos k_F a \sin\theta\sin k_0a|$. However, these gaps are shifted in energy by the shift term $\delta \sin k_Fa \cos\theta \sin ka$ arising from the asymmetry of the dispersion $h_k$. The shifts are equal to $\pm \delta |\sin k_Fa \cos\theta \sin k_0a|$ at the two Fermi points. The system enters a topological superconducting phase only as long as these shifts do not close the gap:
\begin{itemize}
\item At $\theta =0$, i.e., a ferromagnetic arrangement of the impurity spins, the $p$-wave gap vanishes and the system is gapless and nontopological.

\item At $\theta = \pi/2$, i.e., when the spin helix of the impurity spins has zero average magnetization, the shift vanishes and the system always enters a topological superconducting phase as the Shiba bands start to overlap at the chemical potential.   

\item For intermediate $\theta\in(0,\pi/2)$, the system becomes gapless when the shift term becomes larger than the pairing term, i.e., when $|\sin k_Fa\cos\theta|-|\cos k_Fa\sin\theta|>0$. This happens for $\theta<k_Fa<\pi-\theta$ (mod $2\pi$). Thus there are alternating topological and nontopological phases as a function of the Fermi momentum $k_F$ of the superconductor.

\end{itemize}
This scenario is illustrated by the numerical results for the dispersion, the gap function, and the excitation spectrum in Fig.\ \ref{fig:small_xi} and for the phase diagram in Fig.~\ref{fig:small_xi_PD}.

\subsection{Large coherence length}
\label{large}

\begin{figure*}[t]
\begin{centering}
\includegraphics[width=.34\textwidth]{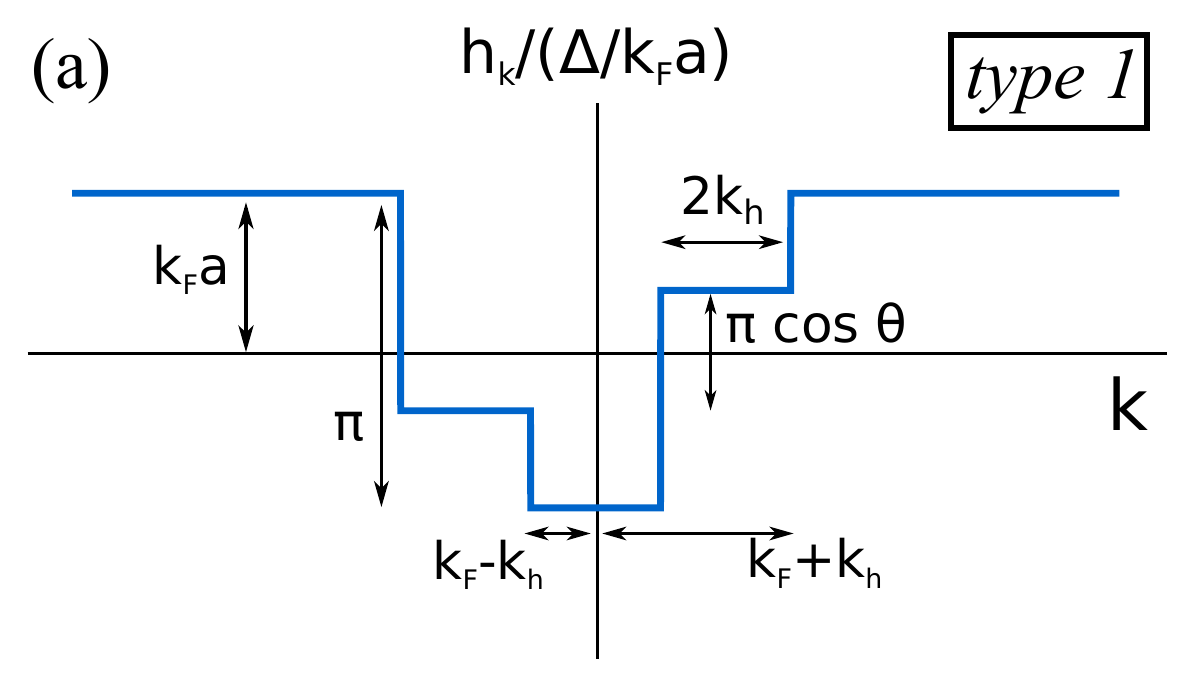}
\includegraphics[width=.34\textwidth]{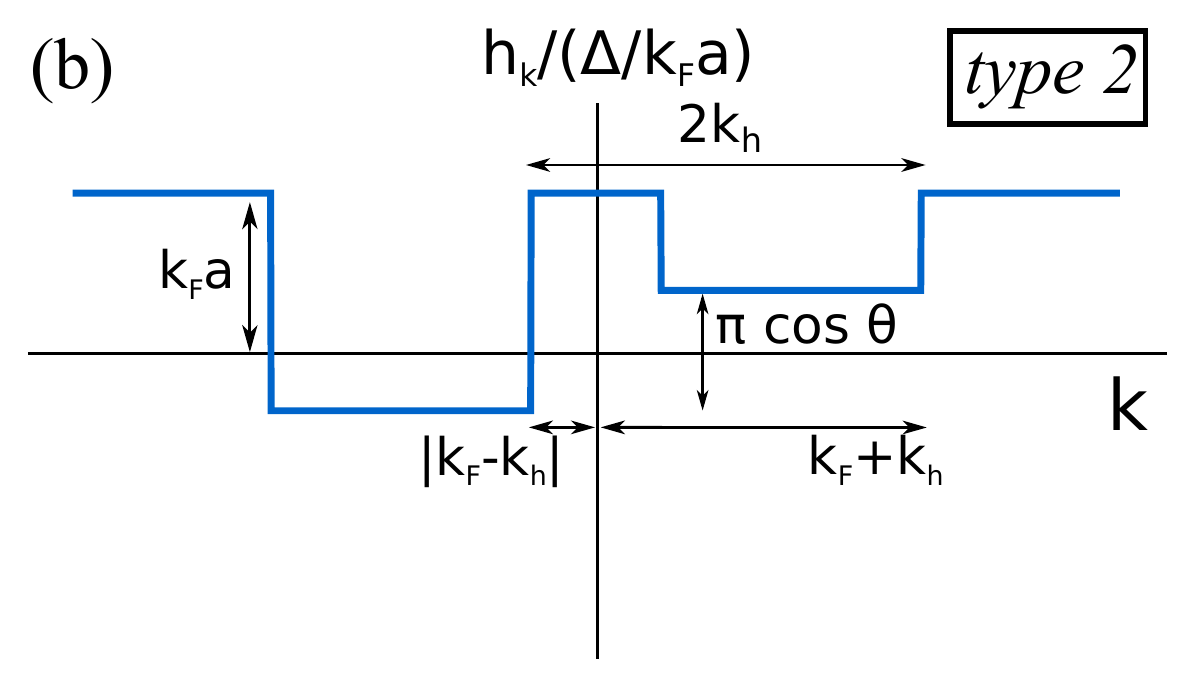}\\
\includegraphics[width=.3\textwidth]{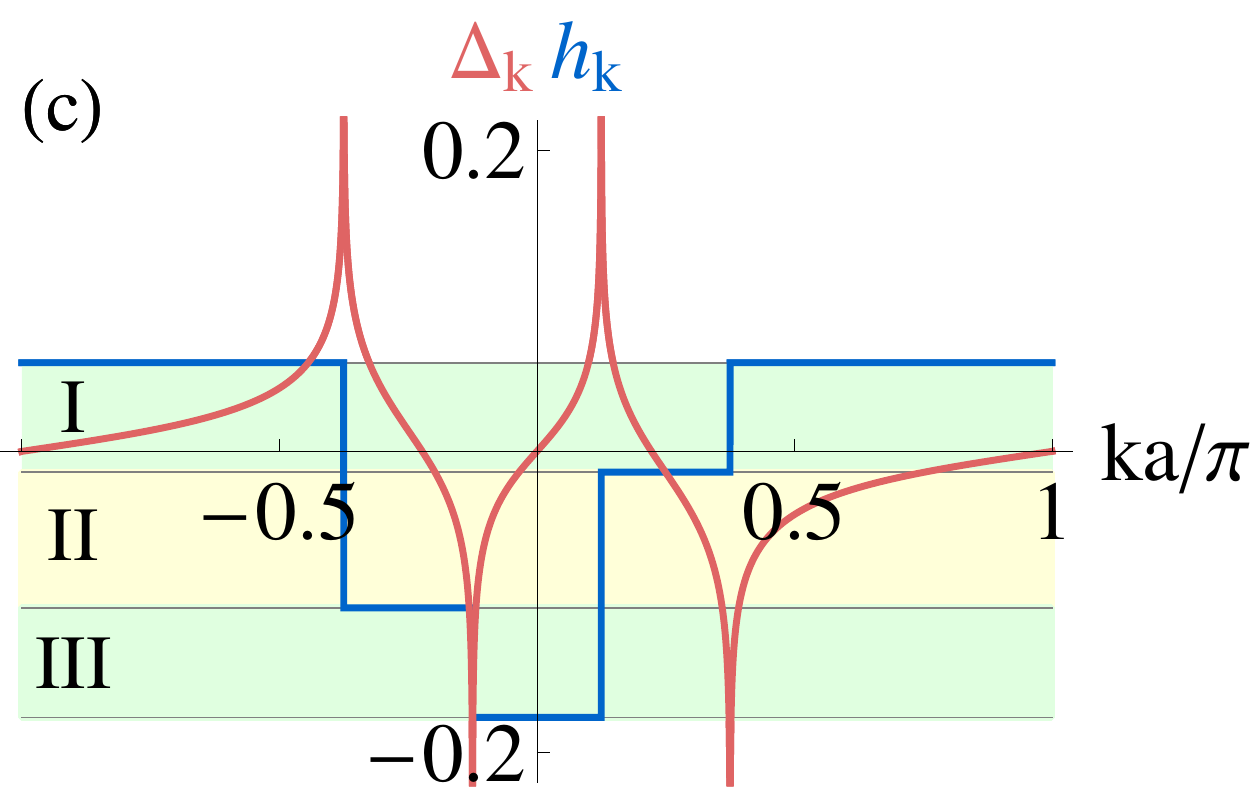}
\includegraphics[width=.3\textwidth]{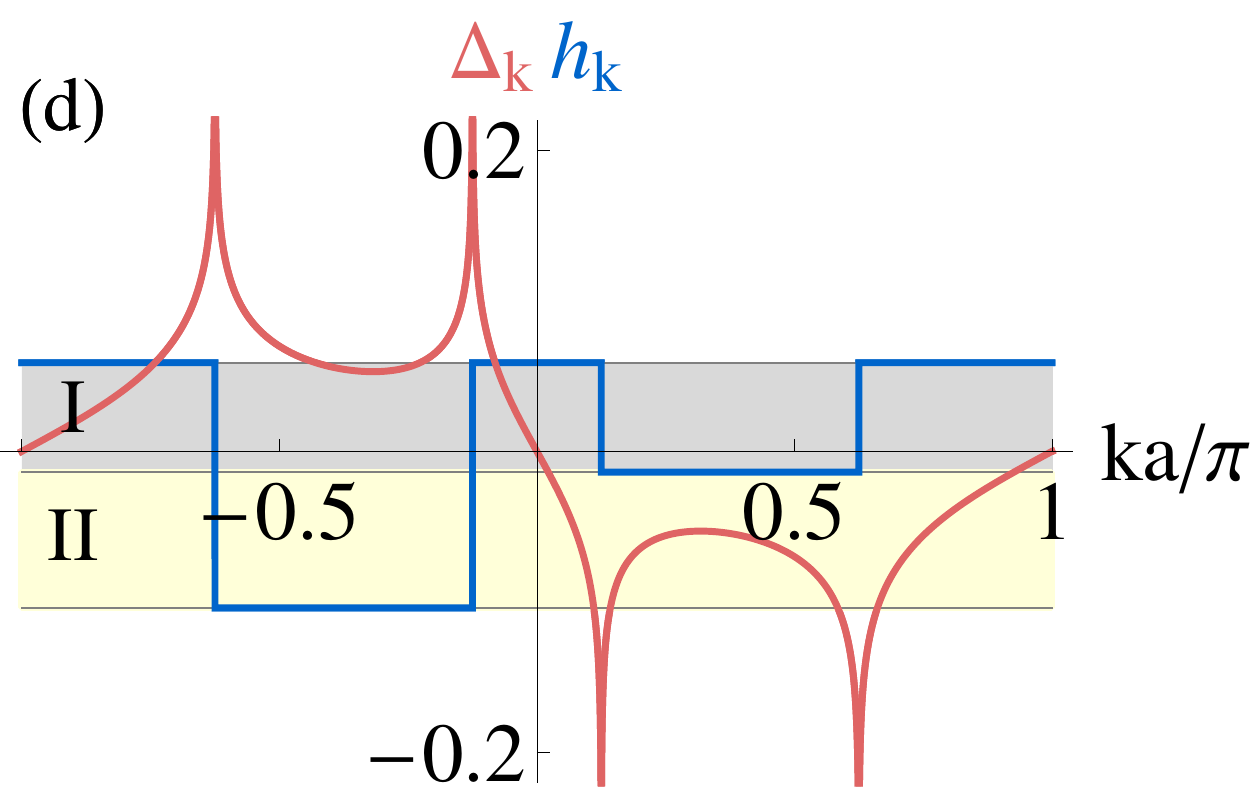}
\par\end{centering}
\caption{(a,b) Schematic plot of the two representative classes of dispersions $h_k$ in the limit of large coherence length $\xi_0\gg a$ as given by the analytical expression in Eqs.~(\ref{hk}) and (\ref{hk-largexi}). In the main text, the two classes are referred to as (a) {\em type 1} and (b) {\em type 2}. The form of the dispersion depends qualitatively on the value of the Fermi and the helix wavevector $k_F$ and $k_h$. (All wavevectors labeling the arrows in (a) and (b) should be understood within the reduced-zone scheme.)
The dispersion is fully symmetric under $k\to -k$ only for $\theta =\pi/2$. (c,d) Dispersions $h_k$ and pairing strengths $\Delta_k$ of both classes for $\epsilon_0=0$, $k_Fa=4.25\pi$, $\theta=3\pi/8$, and (c) $k_ha=\pi/8$, (d) $k_ha=3\pi/8$ (energies are measured in units of $\Delta$).
A nonzero $\epsilon_0$ would lead to an overall shift of the dispersion in energy which causes the chemical potential to pass through various regions as follows: In (c) ({\em type-1} dispersion), there are two regions (I and III, green area) with a symmetric dispersion, for which a topological phase forms. In contrast, in region II (yellow area) $h_k$ is asymmetric and the excitation spectrum $E_k$ becomes gapless. In (d) ({\em type-2} dispersion), $h_k$ has two pairs of symmetric Fermi points in region I (grey area) and the system effectively behaves like a (nontopological) $p$-wave superconducting chain with two channels. In region II, the spectrum may be gapless or trivially gapped. Both classes are shown in the limit of $\xi_0\to\infty$. A large but finite $\xi_0$ would smoothen the jumps in the dispersion and cut off the logarithmic divergences in the pairing strength on the scale of $1/\xi_0$. In addition to these cases, the dispersions $h_k$ can also differ by an overall minus sign, with analogous 
conclusions for the phase diagram.} \label{fig:large_xi}
\end{figure*}

The general Eqs.\ (\ref{hk})  and (\ref{deltak}) for $h_k$ and $\Delta_k$ can also be specified to the limit of large $\xi_0 \gg a$. The dispersion $h_k$ follows from Eq.\ (\ref{hk}) with
\begin{eqnarray}
  F(k) &=& - \left[\arctan \left( \cot \frac{(k_F + k)a}{2}  \right) \right.
  \nonumber \\
 && +\left. \arctan \left(\cot \frac{(k_F - k)a}{2}  \right) \right]  \label{hk-largexi}
\end{eqnarray}
Thus, the dispersion becomes steplike, reflecting the $1/r$ dependence of the hopping amplitudes, with bandwidths of order $\Delta/k_F a$. Depending on the Fermi wavevector $k_F$ and the helix wavevector $k_h$, there are two cases which need to be distinguished. Representative dispersions $h_k$ (referred to as {\em type 1} and {\em type 2} in the following) are shown in Fig.\ \ref{fig:large_xi}.
Note that the sharp steps appear only in the limit $\xi_0\to\infty$. For large but finite $\xi_0$, the steps are smoothened on the scale of $1/\xi_0$. In addition to the dispersions shown in Fig.\ \ref{fig:large_xi}, dispersions of {\em type 1} and {\em type 2} also include the case in which the dispersion differs by an overall minus sign. Then,  the dispersion is of {\em type 1} when $n\pi+k_ha<k_Fa<(n+1)\pi-k_ha$ for some integer $n$ and of {\em type 2} when $n\pi-k_ha<k_Fa<n\pi+k_ha$. 

The pairing strength follows from Eq.\ (\ref{deltak}), where $f(k)$ simplifies to 
\begin{equation} 
  f(k) = -\ln\left[4\sin^2 \frac{ka}{2}\right]
\end{equation}
in the limit $\xi_0\to\infty$. Thus, the paring strength develops logarithmic singularities which occur at those positions where the dispersion $h_k$ develops steps as shown in Fig.\ \ref{fig:large_xi}. Specifically, the pairing strength becomes large and positive near the jumps in $h_k$ which are associated with the $\cos^2 \theta/2$ term in Eq.\ (\ref{hk}) (shown as large jumps in Fig.\ \ref{fig:large_xi}), and large and negative near the jumps which are associated with the $\sin^2 \theta/2$ term in Eq.\ (\ref{hk}). Note that also for the pairing strength, the strict logarithmic divergences are cut off for large but finite $\xi_0$ on a scale of $1/\xi_0$. 

\begin{figure*}[t]
\begin{centering}
\includegraphics[width=.23\textwidth]{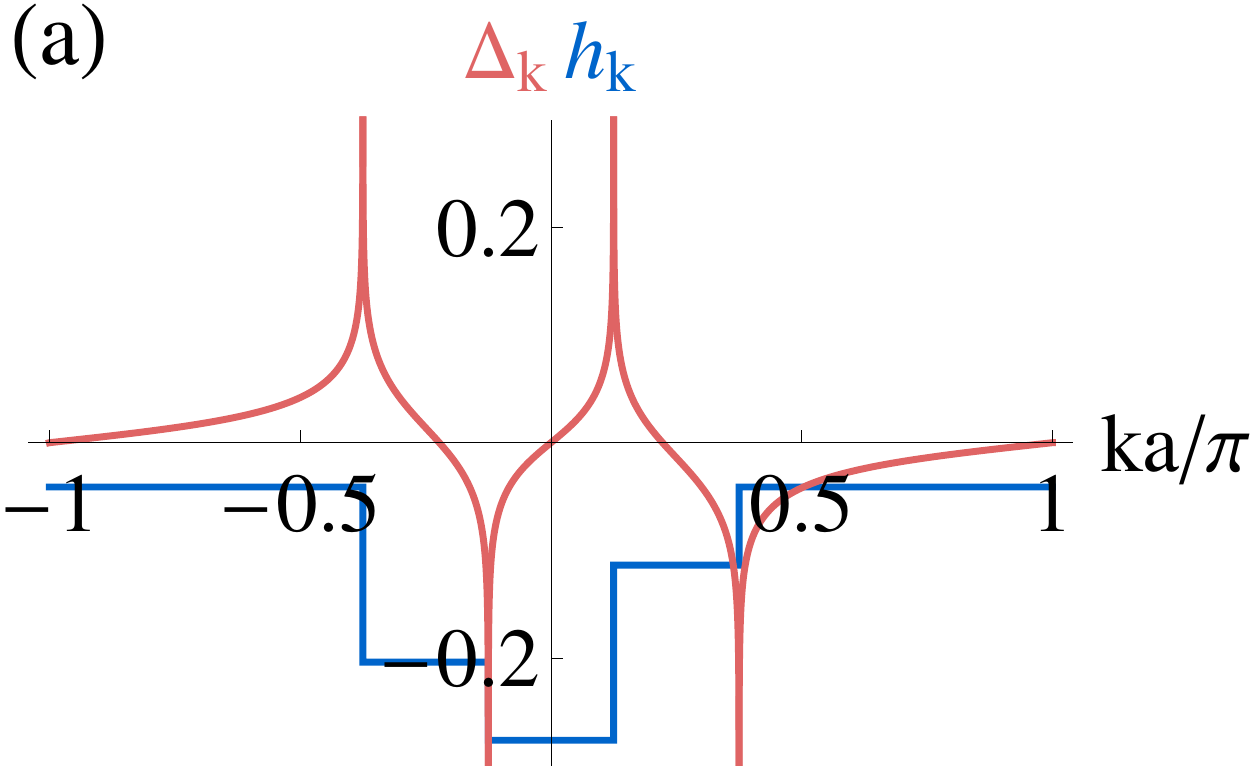} \,\,\,\,\,\,\,\,\,\,
\includegraphics[width=.23\textwidth]{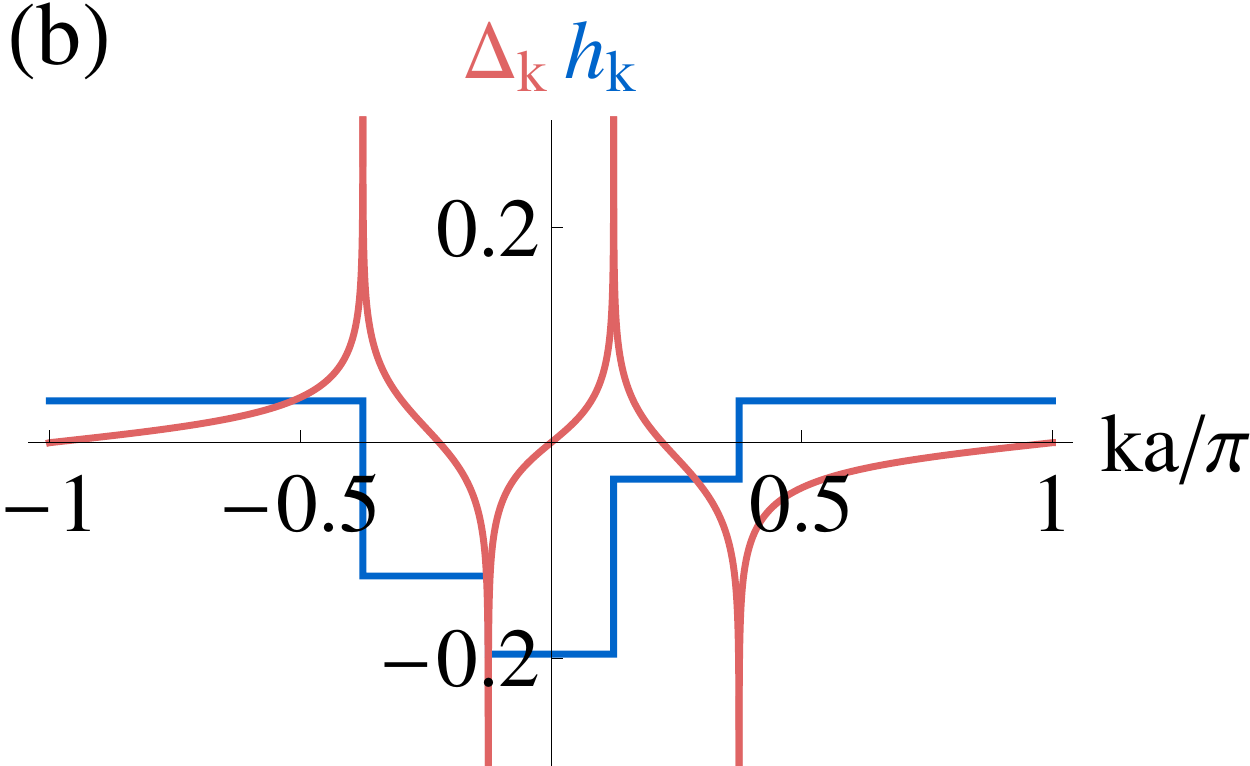} \,\,\,\,\,\,\,\,\,\,
\includegraphics[width=.23\textwidth]{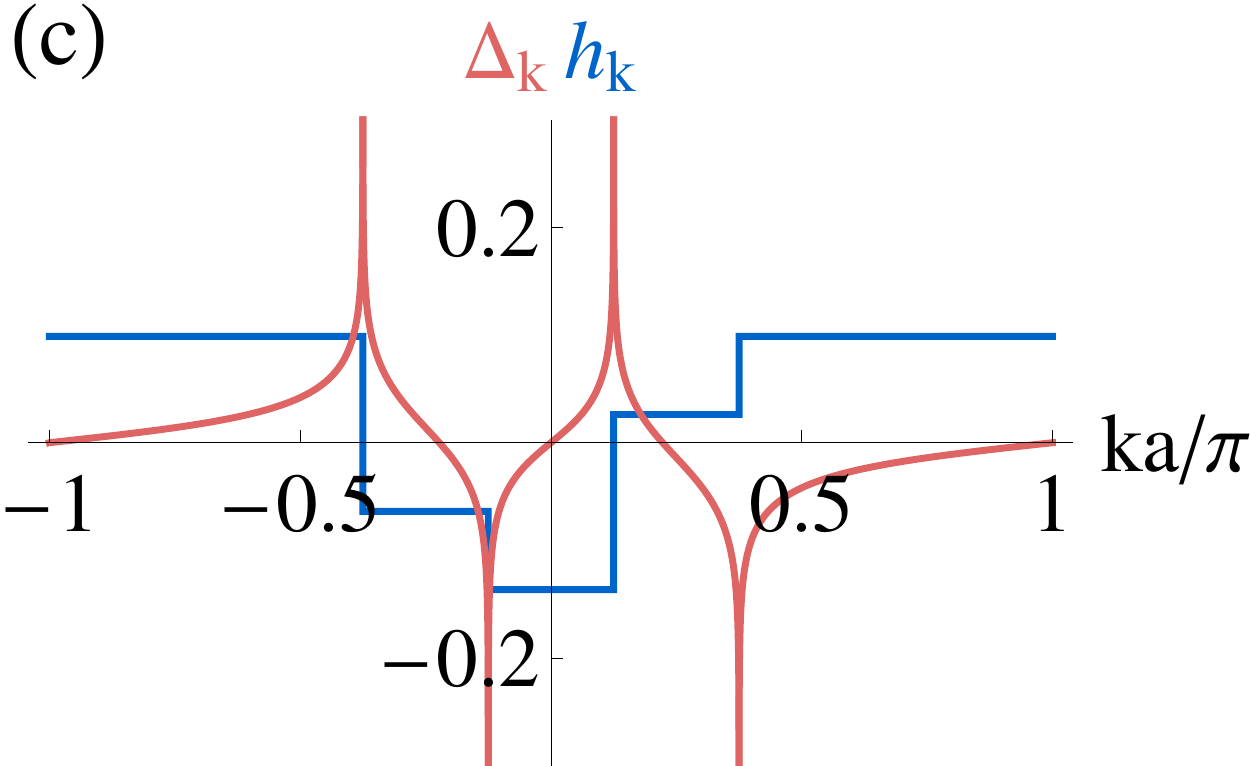}\\
\includegraphics[width=.23\textwidth]{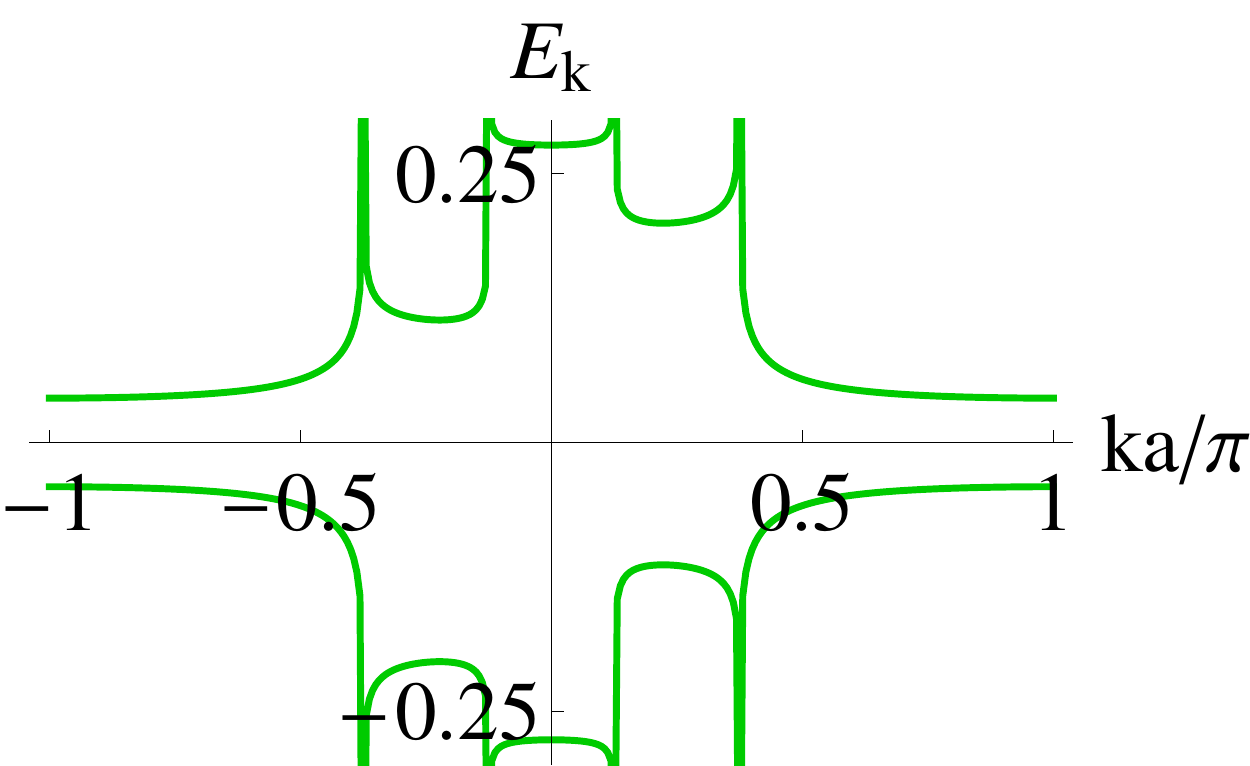} \,\,\,\,\,\,\,\,\,\,
\includegraphics[width=.23\textwidth]{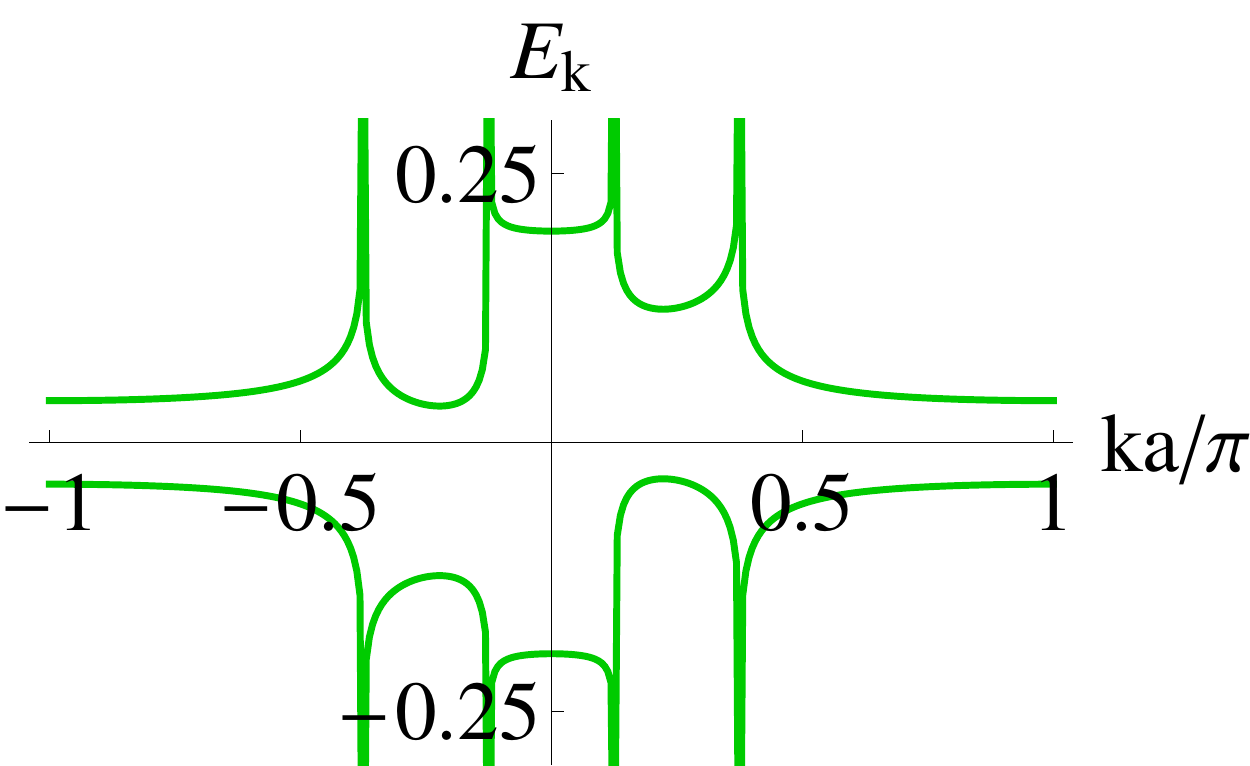} \,\,\,\,\,\,\,\,\,\,
\includegraphics[width=.23\textwidth]{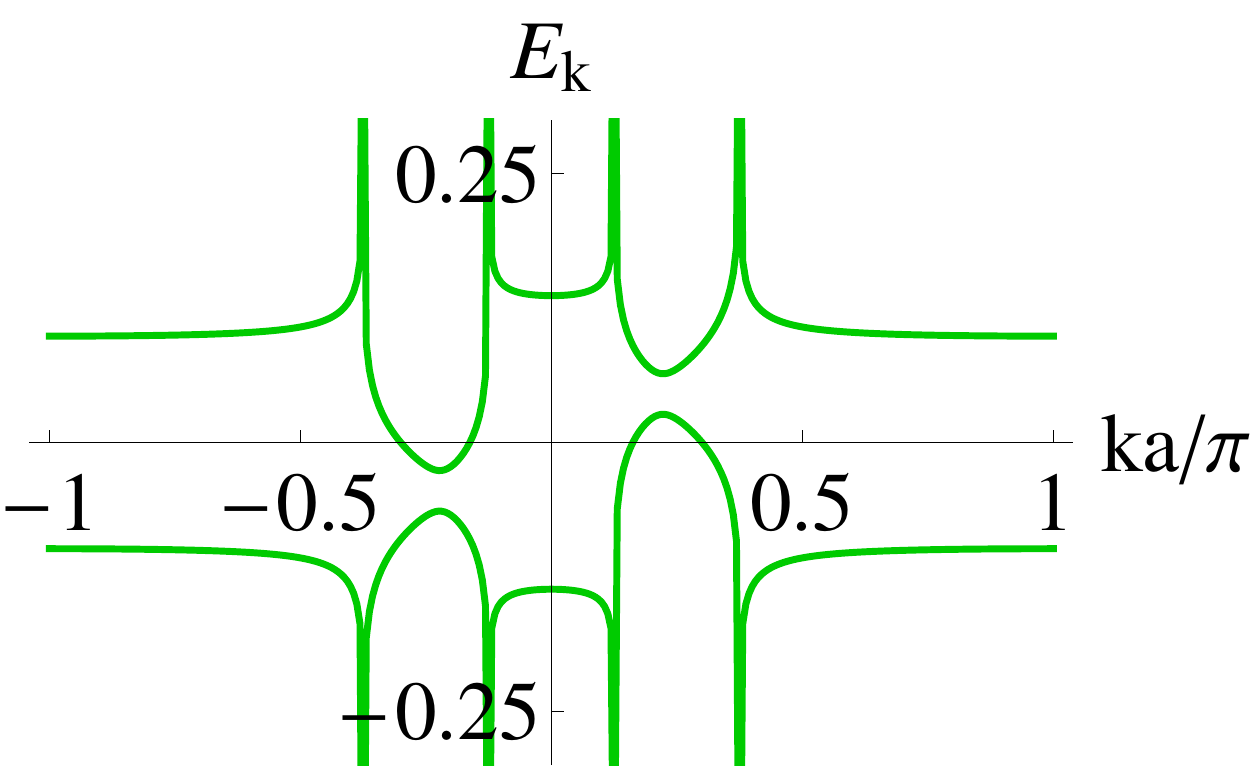}
\par\end{centering}
\caption{{\em Type-1} dispersions $h_k$ and gap functions $\Delta_k$ together with the corresponding excitation spectra [cf.\ Fig.\ \ref{fig:large_xi}(c)]. (a) Chemical potential lies outside the bands of Shiba states ($\epsilon_0=0.1$, gapped nontopological phase). (b) Chemical potential is inside region I ($\epsilon_0=0.02$, gapped topological phase). Analogous results are obtained when the chemical potential is located in region III. (c) Chemical potential is in region II ($\epsilon_0=-0.04$, gapless nontopological phase). For all panels, the remaining parameters are $\xi_0=\infty$, $k_Fa=4\pi+\pi/4$, $\theta=3\pi/8$, and $k_ha=\pi/8$. Energies are measured in units of $\Delta$. 
} \label{fig:option1}
\end{figure*} 

In the following, we discuss the phase diagram separately for dispersions of {\em type 1} and {\em type 2}. To start with, whenever the positive- and negative-energy Shiba bands are nonoverlapping and the chemical potential falls in between the Shiba bands, the system is nontopological, with a large $s$-wave band gap of the host superconductor and a trivial gap between the positive- and negative-energy Shiba bands. The specifics of the dispersion become relevant once the positive- and negative-energy Shiba bands are overlapping (though still well separated from the continuum excitations). 

{\em Type 1.}---As indicated in Fig.\ \ref{fig:large_xi}(c), the dispersion has three characteristic regions. Depending on the energy $\epsilon_0$ of the Shiba states of the individual impurities, the chemical potential (center of the gap of the host superconductor) can be located in any of these regions. In regions I and III, the dispersion $h_k$ is symmetric under $k \to -k$, with one pair of Fermi points. Thus, when the chemical potential falls into these regions, the effective $p$-wave pairing $\Delta_k$ will open a gap at the chemical potential and the Shiba chain enters into a topological superconducting phase. In contrast, when the chemical potential falls into region II, pairing is suppressed by the fact that the dispersion (and hence the two Fermi points) are asymmetric under $k \to -k$. Thus, the overlapping Shiba bands will remain gapless despite the effective $p$-wave pairing $\Delta_k$. In fact, dispersions of {\em type 1} are always gapless in region II. To see this, we again compare the gap to 
the asymmetric energy shift. Since invariably, there are additional zeros of $\Delta_k$ for some pair $\pm k\neq 0,\pi$ [see Fig.~\ref{fig:large_xi}(c)], the shift term always exceeds pairing for particular wavevectors, yielding a gapless spectrum at these points. These findings are illustrated by numerical results for the excitation spectra for various values of $\epsilon_0$, as shown in Fig.\ \ref{fig:option1}.

\begin{figure}[t]
\begin{centering}
\includegraphics[width=.23\textwidth]{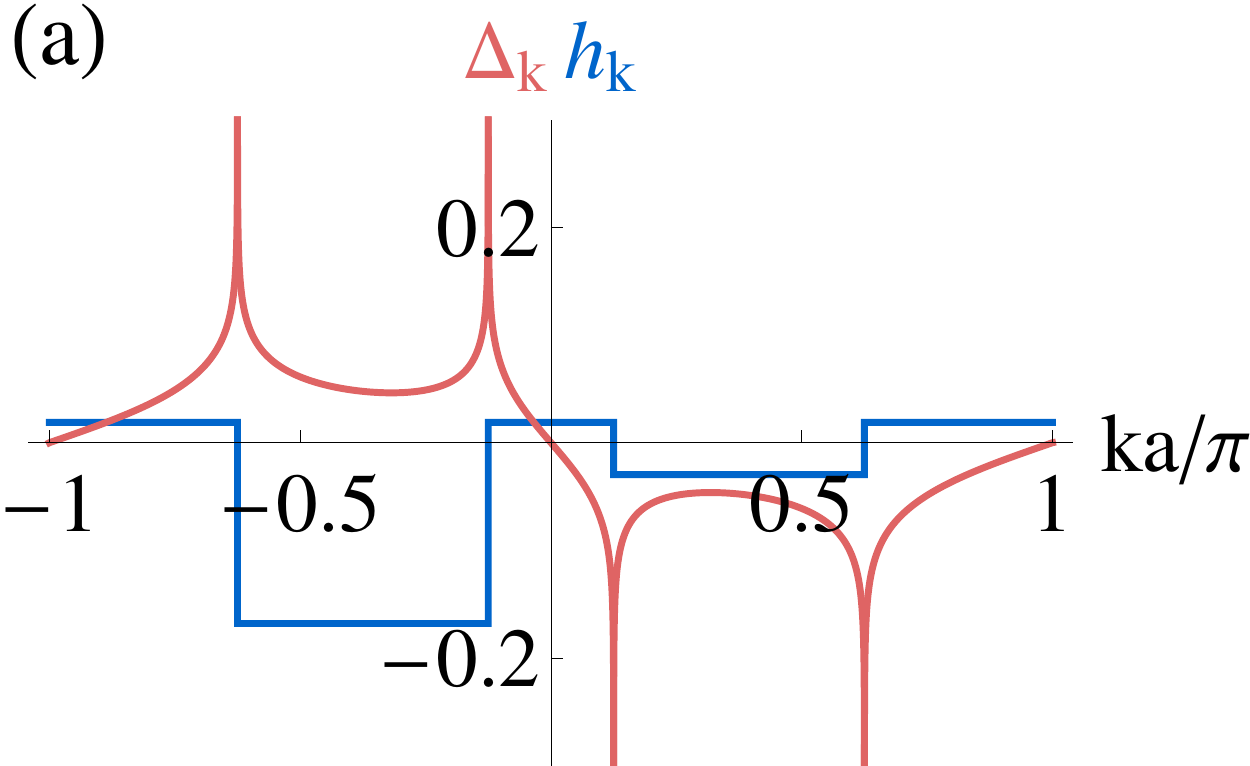}
\includegraphics[width=.23\textwidth]{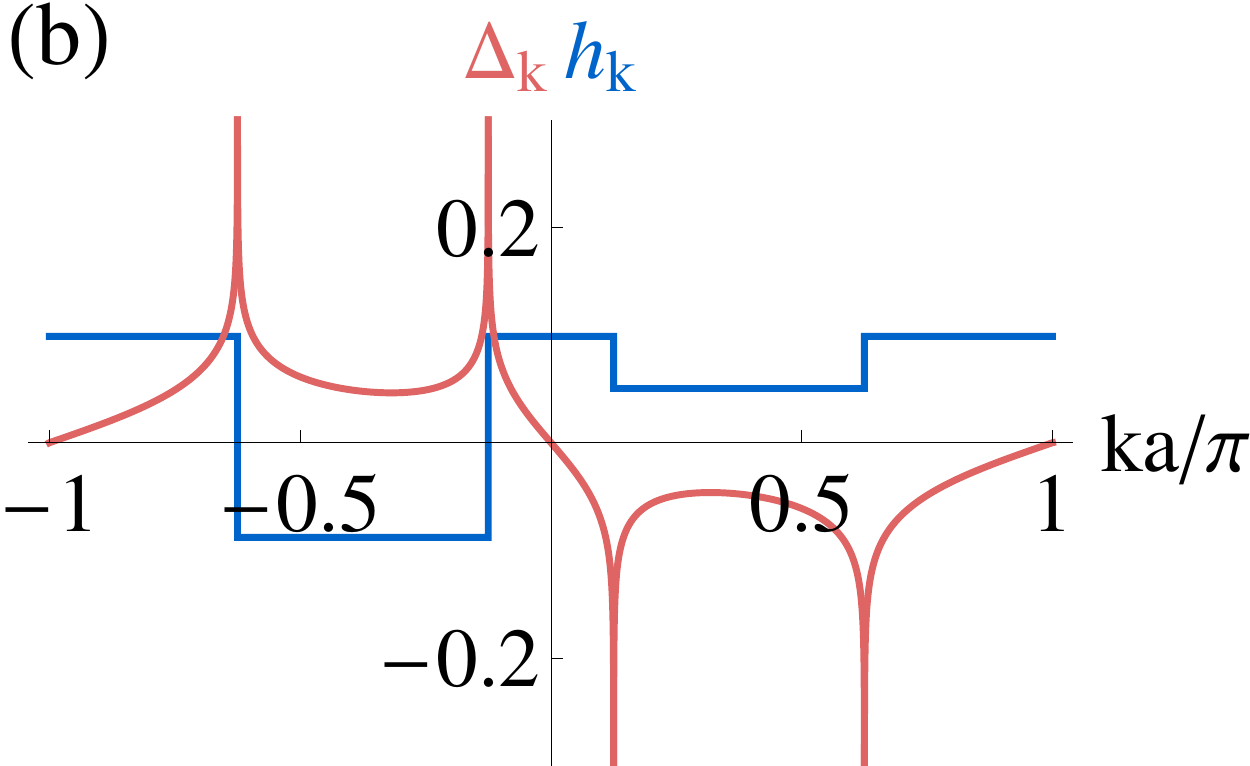}\\
\includegraphics[width=.23\textwidth]{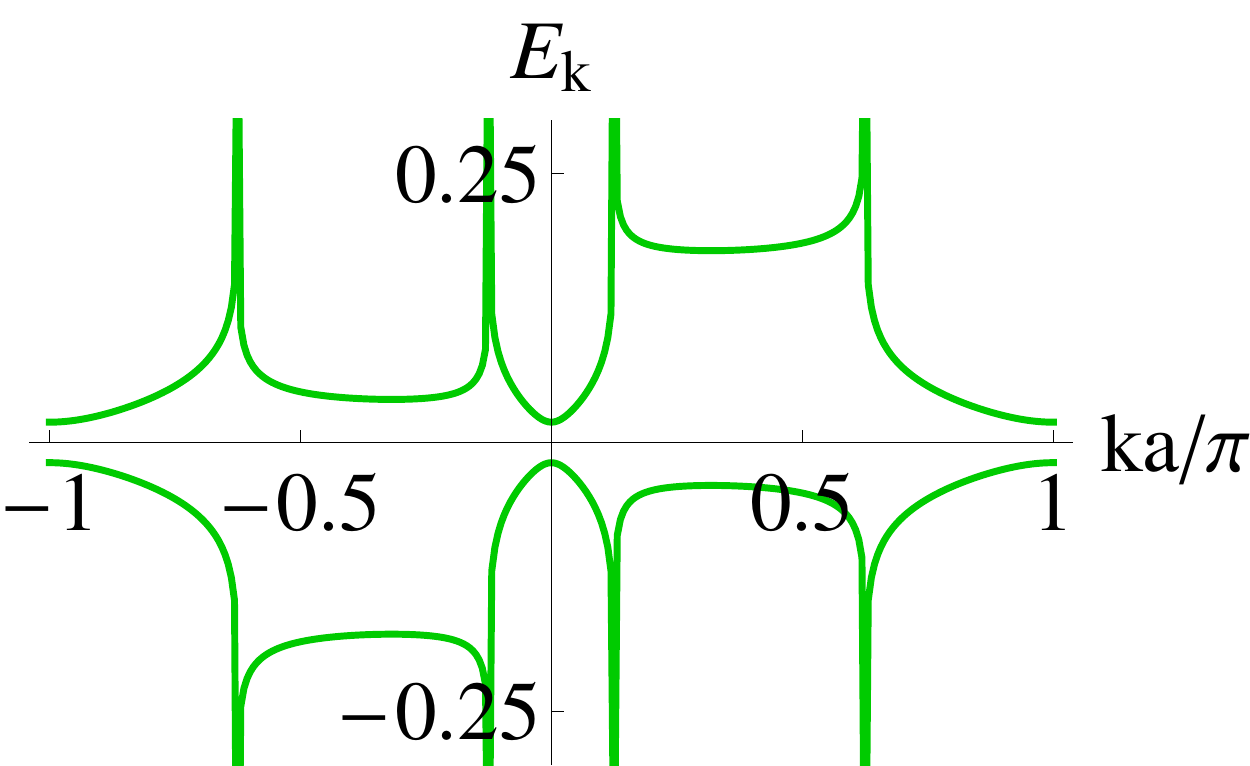}
\includegraphics[width=.23\textwidth]{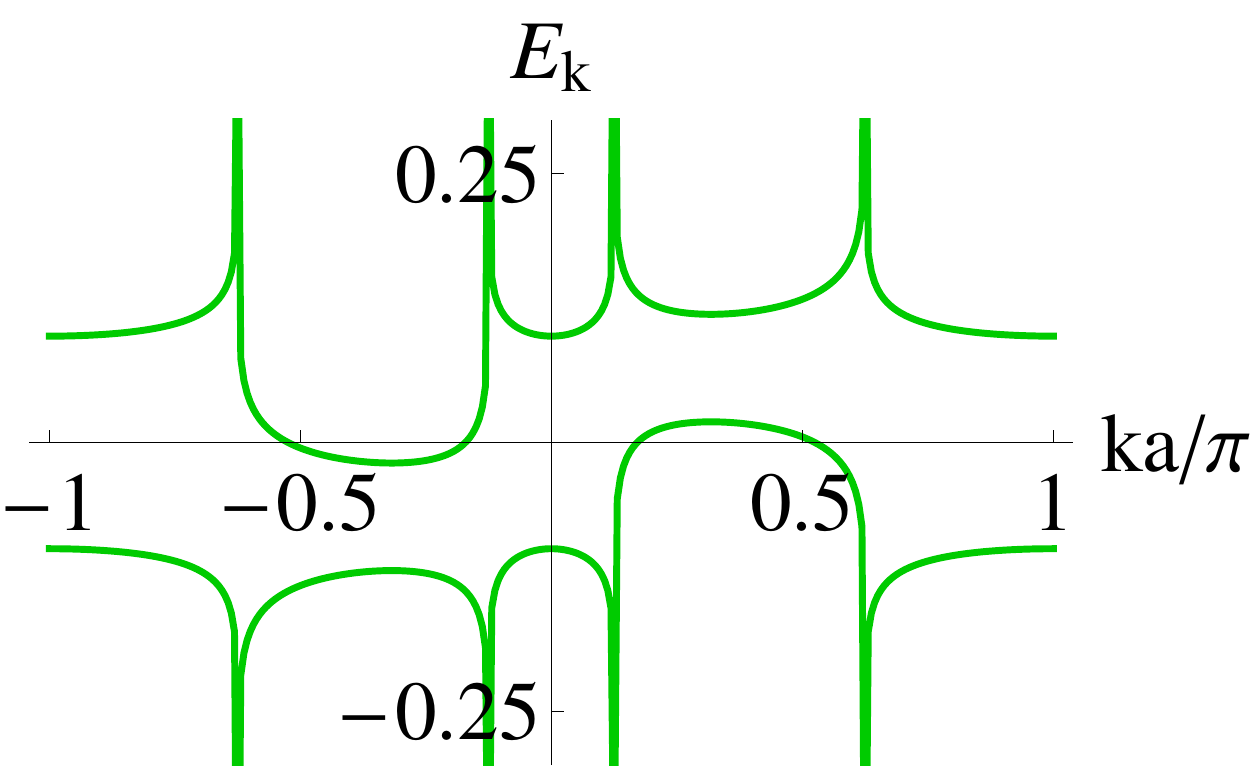}
\par\end{centering}
\caption{Dispersions $h_k$, gap functions $\Delta_k$, and excitation spectra for dispersions of the form shown in Fig.\ \ref{fig:large_xi}(d) ({\em type 2}). (a) Chemical potential lies inside region I in the band of Shiba states (gapped nontopological phase). (b) Chemical potential is in region II (gapless nontopological phase). The parameters are chosen as $\xi_0=\infty$, $k_Fa=4\pi+\pi/4$, $\theta=3\pi/10$, $k_ha=3\pi/8$, and (a) $\epsilon_0=-0.04$, (b) $\epsilon_0=0.04$. Energies are measured in units of $\Delta$.
 }\label{fig:option2}
\end{figure}

{\em Type 2.}---In this case, the dispersion has two characteristic regions as indicated in Fig.\ \ref{fig:large_xi}(d). When the chemical potential falls into region I, there are two symmetric pairs of Fermi points. This is effectively analogous to a two-channel spinless $p$-wave superconducting wire and hence, the system is in a gapped nontopological phase. When the chemical potential falls into region II, the dispersion becomes strongly asymmetric under $k \to -k$. Correspondingly, pairing is suppressed and the system can enter a gapless nontopological phase. In contrast to {\em type-1} dispersions, a dispersion of {\em type 2} can still lead to a gapped (but nontopological) excitation spectrum in region II with asymmetric Fermi points, since $\Delta_k$ has no additional zeros. These conclusions are illustrated by numerical results in Fig.\ \ref{fig:option2}.

\begin{figure}[t]
\begin{centering}
\includegraphics[width=.4\textwidth]{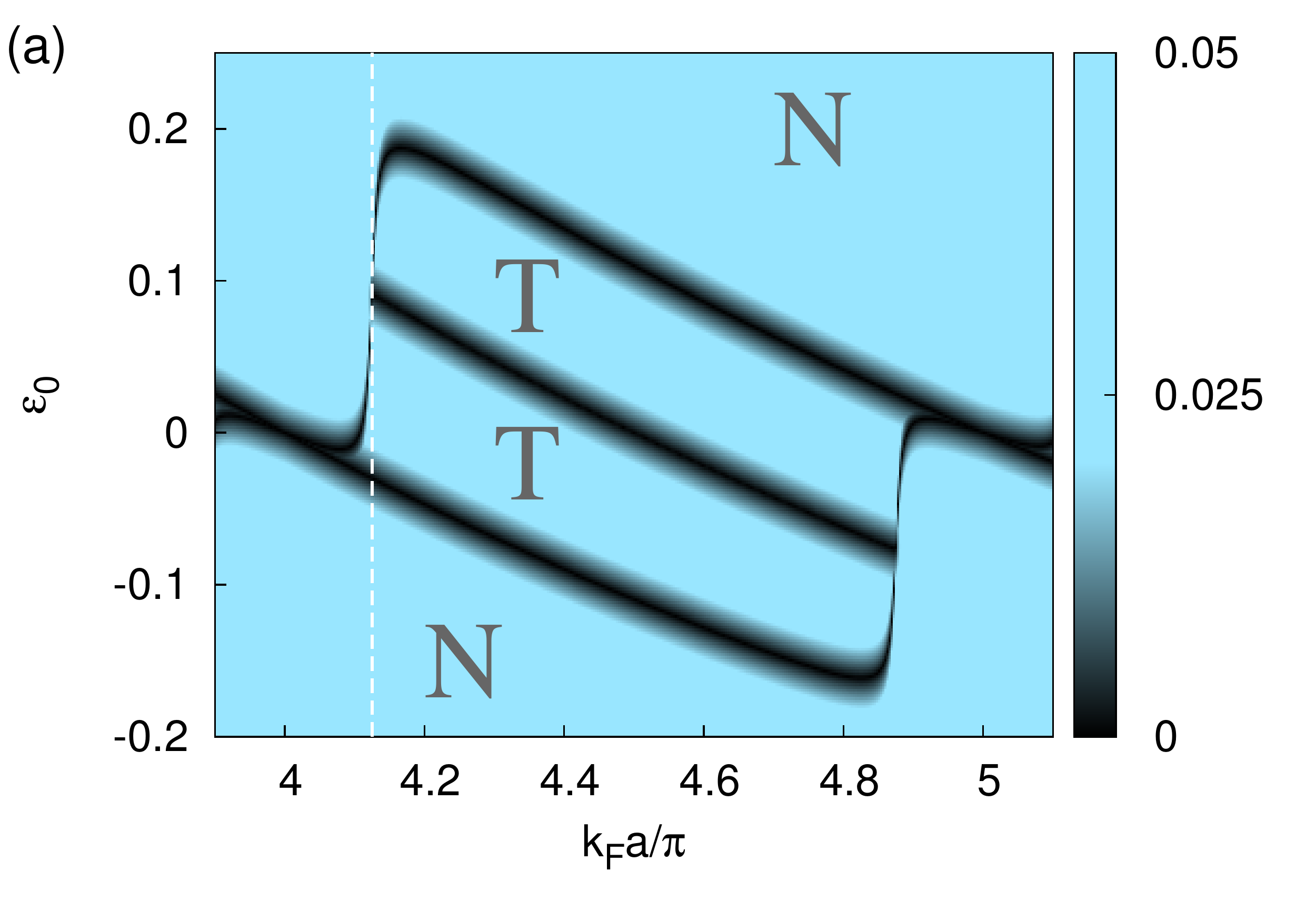}\\
\includegraphics[width=.4\textwidth]{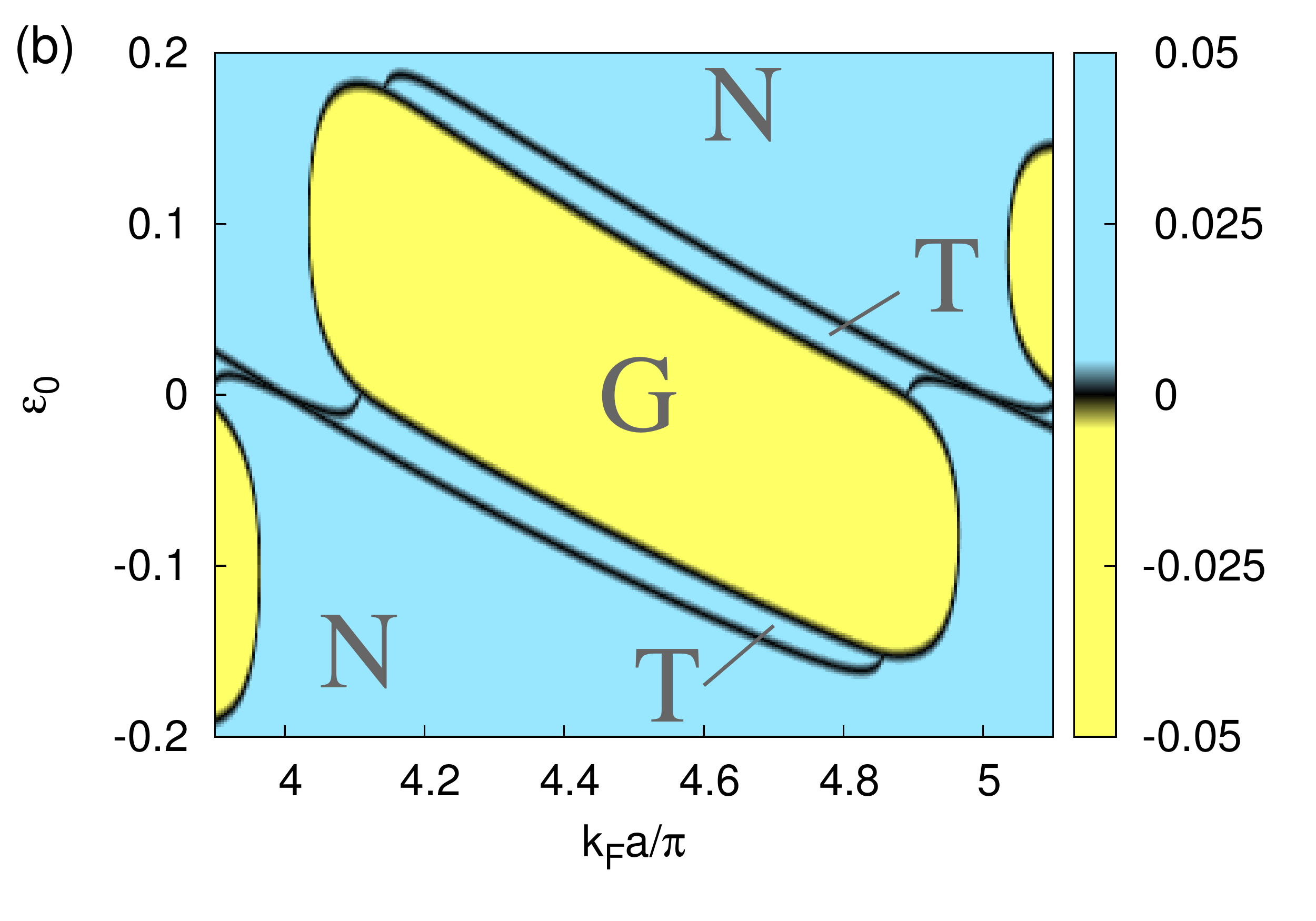}
\par\end{centering}
\caption{Numerical results for the energy minimum of the upper band (color scale) vs. $k_Fa$ and $\epsilon_0$ for a long coherence length $\xi_0=50a$, $k_ha=\pi/8$, $\Delta=1$, and (a) $\theta=\pi/2$, (b) $\theta=\pi/5$. Color scale and labels are as in Fig.~\ref{fig:small_xi_PD}. 
The topological phase transitions at the boundaries of regions I and III in Fig.~\ref{fig:large_xi}(c) appear as diagonal black lines $\epsilon_0=-(k_Fa\ \mathrm{mod}\ 2\pi)\Delta/k_Fa$ in the phase diagram. The almost vertical transition lines between T and N in (a) are associated with the transition between {\em type-1} and {\em type-2} dispersions at $k_Fa = 4\pi+k_ha=4.125\pi$ (white dashed line) and $k_Fa = 5\pi-k_ha=4.875\pi$. As discussed in Sec.~\ref{sec:majorana}, this transition becomes infinitely sharp for $\xi_0\to\infty$. For this reason the gap closing is hardly visible at this numerical resolution in some regions of parameter space. As in the short-$\xi_0$ limit, the topological phase for the symmetric spectrum in (a) is split in half by a metallic line.
At this line, the chemical potential meets the middle plateaux in the dispersion $h_k$ which are at the same height for $\theta=\pi/2$ [see Fig.~\ref{fig:large_xi}(a)]. The excitation spectrum has two simultaneous gap closings at $\pm k_0$. For $\theta<\pi/2$ the spectrum becomes asymmetric and the energy at these two points is shifted in opposite directions. Thus, the metallic line is expanded into the gapless region marked by G in panel (b).
}\label{fig:large_xi_PD}
\end{figure} 

Thus, we find a topological superconducting phase only for dispersions of {\em type 1}. The range of parameters over which the gapped topological phase extends becomes maximal for $\theta = \pi/2$ for which the hopping amplitudes $(h_{\rm eff})_{ij}$ are real, resulting in a dispersion which is symmetric under $k\to -k$. The resulting phase diagram is plotted in Fig.~\ref{fig:large_xi_PD} showing the alternation of topological and gapless phases as well as topological phase transitions as the dispersion changes between {\em type 1} and {\em type 2}. This alternation between topological and nontopological phases is similar to the case of small $\xi_0$. 

For a simple isotropic model of the superconducting host, the RKKY interaction results in a spin helix whose wavevector $k_h$ is directly related to $k_F$, as discussed at the end of Sec.\ \ref{sec:Model}. Remarkably, this relation implies that in this case the system is right at the critical point between {\em type-1} and {\em type-2} dispersions! We will discuss the behavior of the subgap states in the vicinity of this transition in the next section. 

\section{Majorana bound states}
\label{sec:majorana}

Whenever the chain of Shiba states is in the topological phase, one expects localized Majorana bound states to form at the ends of finite chains. In a  semi-infinite chain, the Majorana bound state has strictly zero energy. In a finite wire segment, Majorana bound states form at both ends and overlap in the interior of the wire, thereby acquiring a finite energy splitting. The overlap and hence the energy splitting is controlled by the decay of the Majorana wavefunctions.   

We now turn to a numerical analysis of the decay of the Majorana wavefunction and the corresponding energy splitting for the model given in Eqs.~(\ref{heff}) and (\ref{deltaeff}). In conventional models of 1d topological superconductors such as the Kitaev chain, the Majorana states decay exponentially into the bulk as controlled by the gap of the topological phase. This leads to an energy splitting which is exponentially small in the length of the chain. The Shiba chain differs in that hopping and pairing is long range, exhibiting a power-law decay for distances which are small compared to the coherence length $\xi_0$ of the underlying superconductor. This raises the question of the nature of the decay of the Majorana wavefunctions on scales short compared to $\xi_0$. 

This issue is mute in the case of a short coherence length $\xi_0\ll a$ in which the Shiba chain reduces to a Kitaev chain (with an additional phase gradient for general opening angles $\theta$ of the spin helix). Thus, one expects the conventional behavior in this case and indeed, in the topological phase, this model supports exponentially localized Majorana states whose decay length is determined by the parameters of the Kitaev chain in the usual way. 

\begin{figure*}[tp]
\begin{centering}
\includegraphics[width=.325\textwidth]{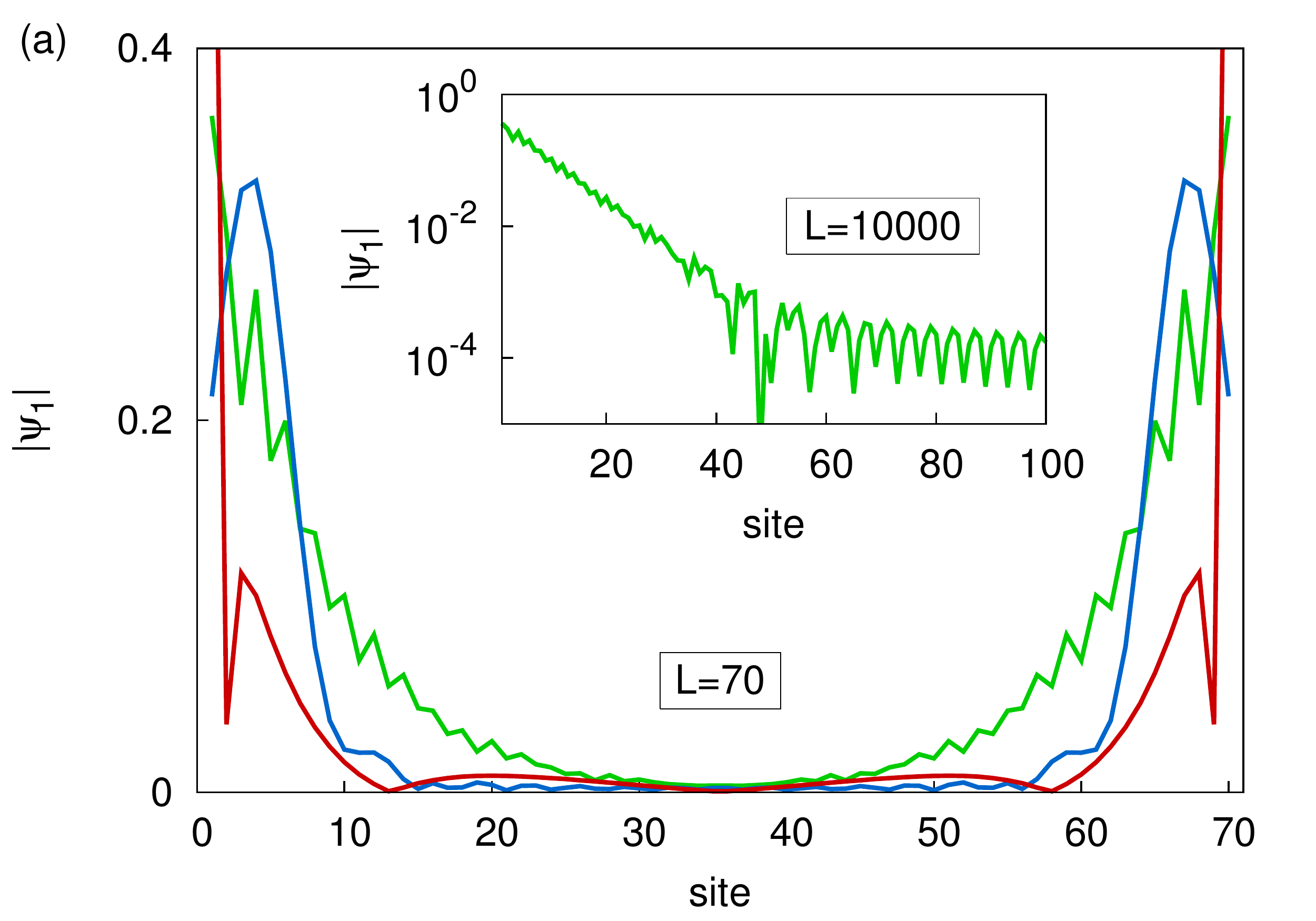}
\includegraphics[width=.325\textwidth]{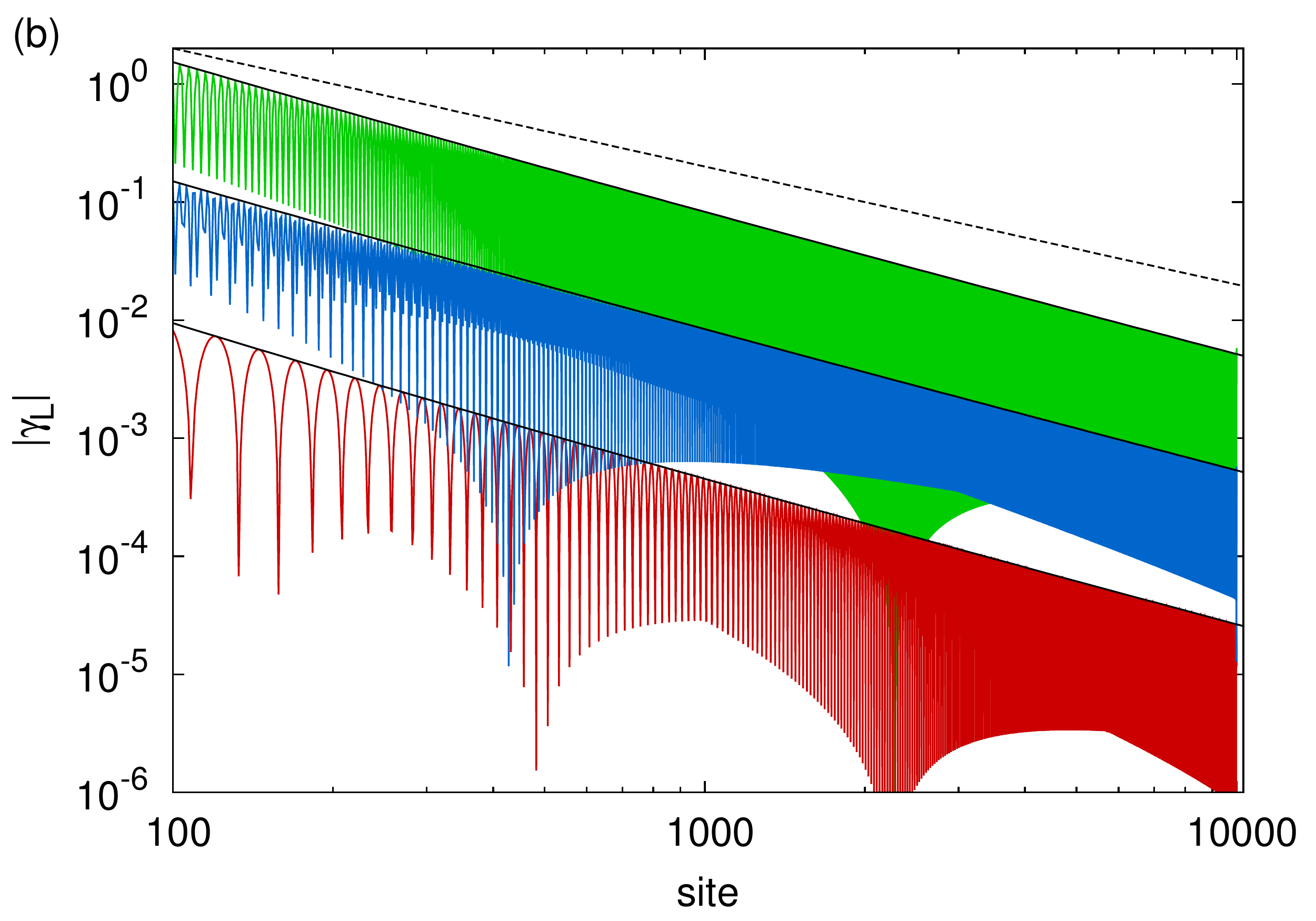}
\includegraphics[width=.325\textwidth]{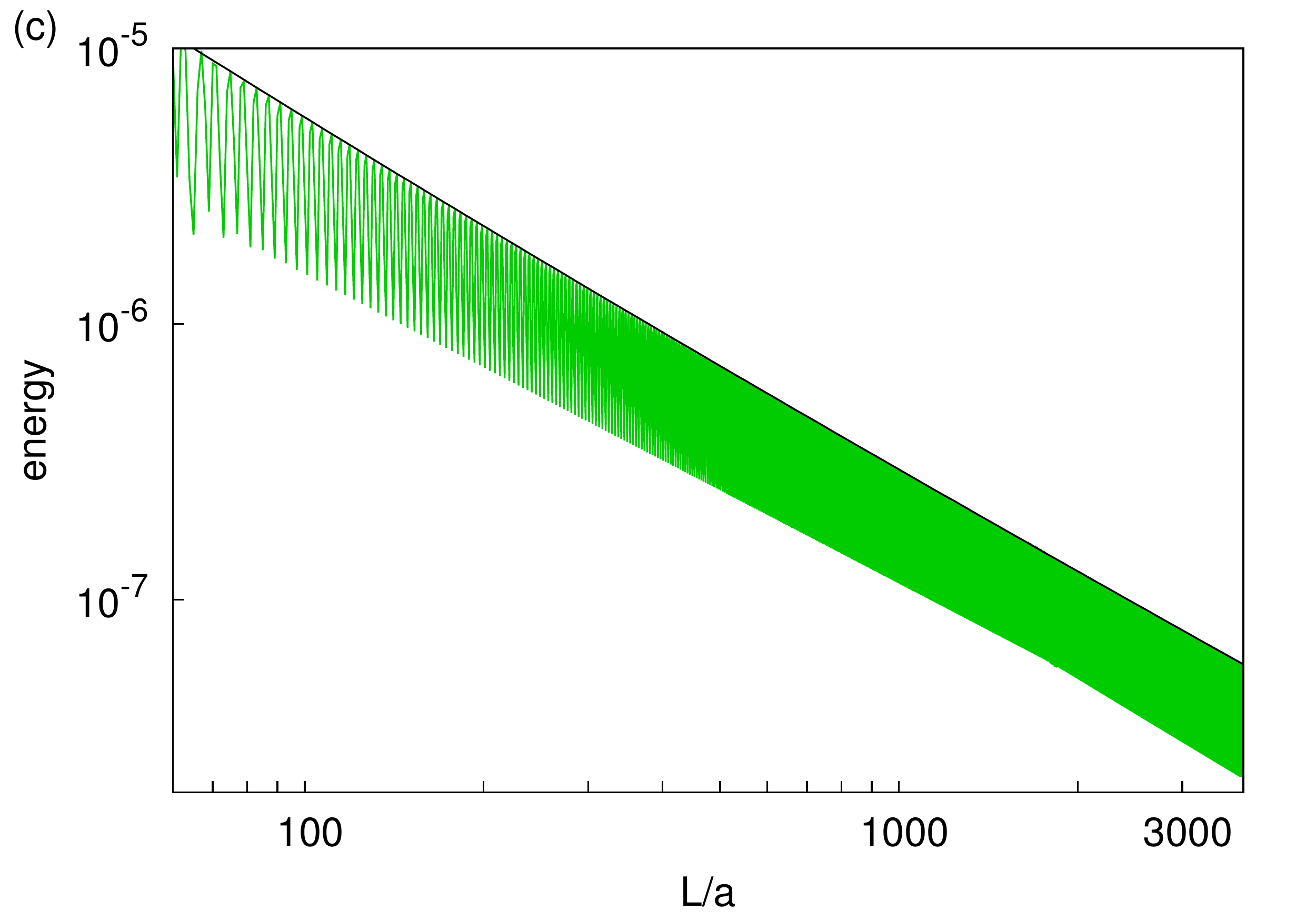}
\par\end{centering}
\caption{(a) Spatial profile of the lowest-energy wavefunction $|\psi_1|=(\phi_1^2+\chi_1^2)^{1/2}$, where $\phi_1$ and $\chi_1$ are the electron and hole components of the Nambu spinor $\psi_1$. All curves are for a chain length $L=70$ and we have set $\xi_0=\infty$, $\Delta=1$, and $\theta=\pi/2$. The remaining parameters are $k_ha/\pi=0.25;0.1;0.26$, $\epsilon_0=-0.01;-0.13;0$, and $k_Fa/\pi=4.5;4.8;4.3$ for the green, blue and red curve, respectively. Inset: Semi-log plot of the first 100 sites of $|\psi_1|$ in chain of length $L=10000$. The parameters are the same as for the green curve in the main panel. The wavefunction initially decays exponentially and then crosses over to a much slower decay. The crossover point depends sensitively on the point in parameter space. (b) Log-log plot of the left Majorana wavefunction $|\gamma_L|$ for a chain of length $L=10000$. (The first 100 sites are not shown.) The three curves are for the same set of parameters as in (a) and shifted vertically for clarity. The 
black solid lines represent $1/[x\ln^2(x/x_0)]$ fits to the envelopes of the curves. The dashed lines shows a $1/x$ power law for comparison. The Majorana wavefunctions can be obtained from the lowest energy wavefunction by a rotation in Nambu space\protect\cite{footnote1} $\gamma_{L/R}=\chi_1\pm i\phi_1$. The obtained fit parameters are $x_0/a\sim 0.17,0.30,0.55$ for the three curves. (c) Log-log plot of the Majorana energy splitting {\em vs} chain length for the same parameters as the green curve in (a) and (b). Similar to the wavefunction decay, the envelope of the energy splitting fits a $1/[x(\ln(x/x_0))^2]$ law (black line) with $x_0/a\sim 0.22$. }\label{fig:decay1}
\end{figure*} 

\subsection{{\em Type-1 dispersions}}

We now turn to the more interesting (and more realistic) case of large coherence lengths, $\xi_0\gg a$, and parameters such that the dispersion is of {\em type 1}. Also in this case, the Majorana bound state decays exponentially on scales large compared to $\xi_0$. Indeed, given the effective bandwidth of the Shiba bands of order $\Delta/k_Fa$ [cf.\ Eqs.~(\ref{heff})] and the smoothing of the steps in $h_k$ of order $1/\xi_0$, we have an effective Fermi velocity of order $\Delta \xi_0/k_Fa$. Combining this with the effective strength of the $p$-wave pairing of order $\Delta/k_Fa$ [cf.\ Eqs.~(\ref{deltaeff})], we find that the characteristic length scale (analogous to the relation $\xi_0 = \hbar v_F/\Delta$) is indeed of order $\xi_0$. Note that this is only a rough order-of-magnitude estimate which neglects the dependence on the opening angle $\theta$, the energy $\epsilon_0$ of the Shiba bound states, etc. 

The decay of the Majorana bound states on length scales shorter than $\xi_0$ can be readily investigated numerically. To do so, we take the model defined by Eqs.~(\ref{heff}) and (\ref{deltaeff}) and formally set $\xi_0=\infty$. As one readily checks numerically, the resulting model correctly reproduces the behavior of the Majorana bound states of the more complete model with finite $\xi_0$ on scales smaller than $\xi_0$. Numerical results for the $\xi_0 = \infty$ model and for parameters such that the system is in the topological phase are shown in Fig.\ \ref{fig:decay1}. We find that asymptotically, the Majorana bound state decays approximately as a power law with logarithmic corrections. Indeed, the envelope of the Majorana wavefunction can be fit quite accurately by a decay of the type $1/[x\ln^2 (x/x_0)]$ for a variety of parameter sets, cf.\ Fig.\ \ref{fig:decay1}. While similar, this decay is faster than the decay of the hopping and pairing amplitudes. Interestingly, this implies that in a finite wire 
of length $L\ll \xi_0$ (as may well be the case in an experiment) the energy splitting of the two Majorana bound states is not exponentially but merely power-law suppressed in $L$ (with logarithmic corrections). This is illustrated numerically in Fig.\ \ref{fig:decay1}(c).

\subsection{{\em Type-2 dispersions}}

\begin{figure}[t]
\begin{centering}
\includegraphics[width=.46\textwidth]{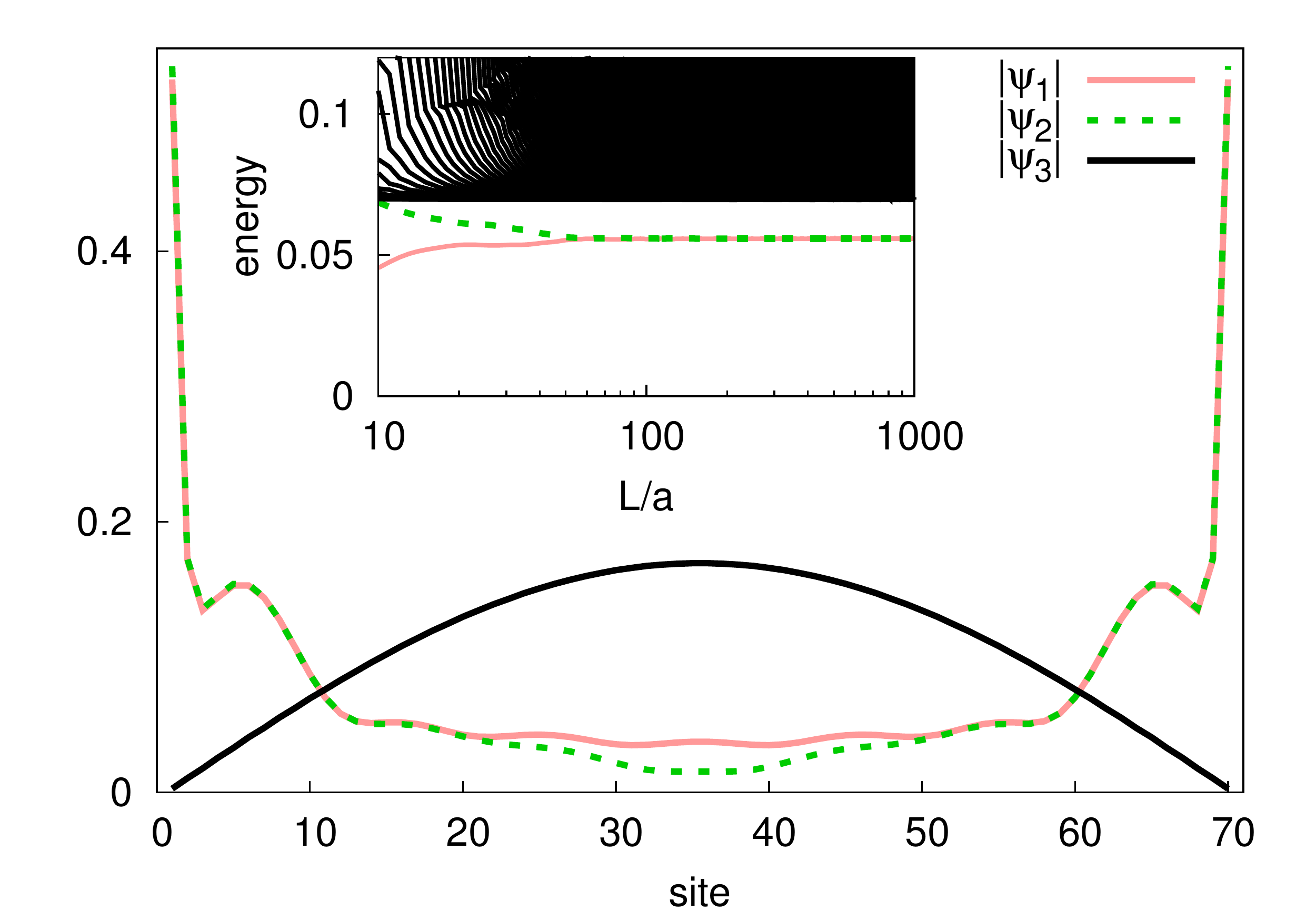}
\par\end{centering}
\caption{Spatial wavefunction profile $|\psi|$ of the first three positive-energy states of a chain with 70 sites with a {\em type-2} dispersion in region I [see Fig.~\ref{fig:large_xi}(d)]. There are two states localized at the ends of the chain. Inset: excitation spectrum of a finite chain as a function of chain length $L$. The plot shows that the two end states remain at a nonzero subgap energy for large $L$. This is the expected behavior of a two-channel $p$-wave superconducting chain with two coupled Majorana bound states at each end. The third state is a bulk state which defines the edge of the quasiparticle continuum. The parameters are: $\Delta=1$, $\epsilon_0=0.05$, $\theta=\pi/2$, $k_ha=\pi/8$, $k_Fa=4.08\pi$.}
\label{fig:en-MBS}
\end{figure}

In the discussion of the phase diagram for large coherence length in Sec.~\ref{large}, we showed that there is no topological phase for dispersions of {\em type 2}. Nevertheless, when the chemical potential is in region I [see Fig.~\ref{fig:large_xi}(d)], the spectrum is analogous to that of a two-channel $p$-wave superconductor. Each of the channels individually supports one Majorana bound state at each end. The hard-wall boundary introduces scattering between the two channels and the Majoranas acquire a finite energy splitting, which is usually of the order of but smaller than the gap.
Numerically, we indeed find two positive-energy subgap states in this regime, one for each end of the chain, as seen in Fig.\ \ref{fig:en-MBS}. Their energy as a function of length is shown in the inset. The two states are clearly separated from the continuum, but they remain at a finite energy even at very large chain lengths. The energy depends on the boundary-induced coupling of the two channels. For long chains, the two subgap states become degenerate as the two ends of the chain are decoupled.

\subsection{Vicinity of the transition between type-1 and type-2 dispersions}

We now turn to a closer look at the transition point $k_Fa = k_ha +n\pi$ and its immediate vicinity, as this is the relevant point of the phase diagram for sufficiently simple host superconductors. When the chemical potential is in region I of a {\em type-2} dispersion and the dispersion changes from {\em type 2} to {\em type 1}, say by tuning $k_Fa$ across $k_ha+n\pi$, the system undergoes a phase transition from an effective two-channel to a single-channel $p$-wave superconducting chain. For $\xi_0=\infty$, this transition is abrupt and the bulk gap does not approach zero on either side of the transition. This can be seen by comparing Figs.~\ref{fig:large_xi}(a) and (b). The transition from {\em type 1} to {\em type 2} involves shrinking, flipping, and extending again the central plateau in $h_k$ near $k=0$. This rather abrupt behavior is manifested in the color-scale plot of the bulk gap in Fig.~\ref{fig:large_xi_PD}(a), where the vertical transition lines become very fine indicating a sharp transition. 
For finite $\xi_0$, the bulk gap closes smoothly as $|k_Fa-(k_ha+n\pi)|\to 0$ on a scale of $1/\xi_0$. 

\begin{figure}[t]
\begin{centering}
\includegraphics[width=.46\textwidth]{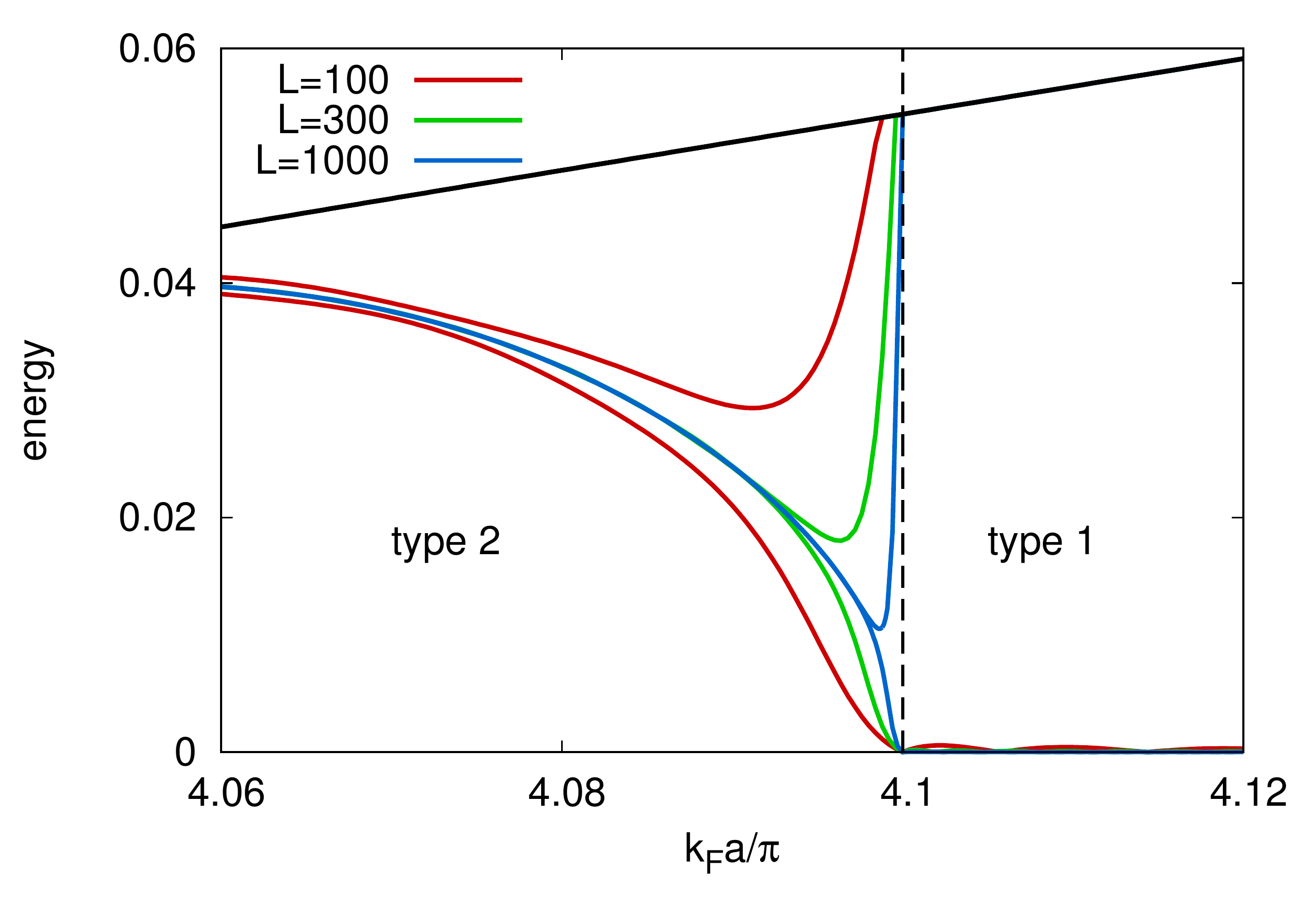}
\par\end{centering}
\caption{Spectrum of subgap states (limited to positive energies in units of $\Delta$) {\em vs} Fermi wavevector $k_F$ near the phase transition from {\em type 2} to {\em type 1} at $k_Fa=4.1\pi$ (dashed line). The plot is for $\xi_0\to\infty$ and chemical potential in region I so that at the transition, the system changes from a two-channel to a single-channel $p$-wave superconductor. The parameters are chosen as $k_ha=0.1\pi$, $\xi_0=\infty$, and $\epsilon_0=0.03$. The colored lines represent the two subgap states for various chain lengths (see legend) and the black line marks the lowest continuum excitation (which is indistinguishable for the different chain lengths). Just before the phase transition on the {\em type-2} side, the two subgap states split. While one state is absorbed into the continuum at the transition, the second state drops to near-zero energies and becomes a Majorana bound state.}
\label{fig:transition}
\end{figure}

The behavior of the subgap spectrum in the vicinity of the transition is illustrated numerically in Fig.\ \ref{fig:transition}. As the transition is approached from the side with the {\em type-2} dispersion (two-channel side), the coupling of the two channels and hence the energy splitting of the end states depends on the difference of the Fermi momenta $\pm (k_F-n\pi)\pm k_h$ of the channels. The inner pair of Fermi momenta approaches zero continuously at the transition and the coupling between the two channels becomes weak. Simultaneously, the closing of one channel is associated with a delocalization of one of the Majorana modes, such that at the transition, the corresponding end state merges with the continuum spectrum. In contrast, the second end state becomes a {\em bona fide} zero-energy Majorana bound state. Interestingly, for finite-length chains, Fig.\ \ref{fig:transition} shows that right at the transition point, there is just a {\em single} almost-zero-energy subgap state. Thus, 
phenomenologically, the chain effectively behaves as if it was in the topological phase. 

\section{Conclusions}
\label{sec:conclusions}

Chains of magnetic impurities placed on a conventional $s$-wave superconductor constitute a promising venue for Majorana physics. Assuming that the magnetic impurities form a spin helix, the bands of Shiba states formed in the host superconductor can enter into a topological superconducting phase. In this paper, we considered the limit of dilute impurities inducing deep Shiba states in which the bands of Shiba states do not overlap with the quasiparticle continuum. Starting with the individual Shiba states, we derived an effective tight-binding Bogoliubov-de Gennes Hamiltonian. While this Hamiltonian has close similarities with the Kitaev model, it differs in important ways: (i) Both hopping and pairing are long range, falling off as $1/r$ for distances small compared to the superconducting coherence length $\xi_0$. (ii) For generic spin helices, the hopping amplitudes are complex (or, equivalently, the pairing amplitude involves a spatially varying phase). These differences have significant consequences, 
both for the phase diagram and for the decay (and hence the energy splitting) of Majorana bound states. Most importantly, the long-range nature of hopping and pairing implies that over a wide range of length scales, the spatial decay of the Majorana bound states is well fit by a power law with logarithmic corrections rather than an exponential dependence. Moreover, the complex hopping amplitudes tend to suppress topological superconductivity; they result in asymmetric dispersions under momentum reversal which suppresses Cooper pairing. As a result, fully planar spin helices whose additional reflection symmetry results in purely real hopping amplitudes, are optimal for realizing topological superconductivity. 

Our approach can be extended in several directions. For instance, it might be relevant for experiment to include spin-orbit coupling within the superconducting host or to extend our approach based on Shiba states to the limit of shallow impurities. 

\begin{acknowledgments}
We thank Norman Yao and Ali Yazdani for stimulating discussions and acknowledge financial support by the Helmholtz Virtual Institute ``New states of matter and their excitations,'' SPP1285 of the Deutsche Forschungsgemeinschaft, the Studienstiftung d.\ dt.\ Volkes, and NSF DMR Grants 0906498 and 1206612. We are grateful to the Aspen Center for Physics, supported by NSF Grant No.\ PHYS-106629, for hospitality while this work was initiated.
\end{acknowledgments}

\appendix

\section{Some integrals}
\label{integrals}

In this Appendix, we evaluate and discuss the integral 
\begin{equation}
I = \int \frac{d{\bf p}}{(2\pi)^3} \frac{e^{i{\bf p}{\bf r}}}{E - \xi_{\bf p} \tau_z - \Delta \tau_x},
\end{equation}
which is used in Secs.\ \ref{sec:shiba_single} and \ref{sec:shiba_chain}. Note that we are interested in subgap energies $E < \Delta$. Explicitly inverting the matrix and changing integration variables to $\xi_p$ and $x=\cos\theta_{\bf p}$ with the polar angle $\theta_{\bf p}$ measured relative to ${\bf r}$, we have
\begin{equation}
  I = \frac{\nu_0}{2} \int d\xi_p \int_{-1}^1 dx  {e^{iprx}} \frac{E + \xi_{\bf p} \tau_z + \Delta \tau_x}{E^2 - \xi_{\bf p}^2 - \Delta^2 }.
\end{equation}
Here, $\nu_0$ denotes the normal-state density of states per spin direction of the superconductor. Thus, we need to evaluate the integrals
\begin{eqnarray}
  I_0 &=& \frac{\nu_0}{2} \int d\xi_p \int_{-1}^1 dx  \frac{e^{iprx}}{E^2 - \xi_{\bf p}^2 - \Delta^2 }, \\
  I_1 &=& \frac{\nu_0}{2} \int d\xi_p \int_{-1}^1 dx \frac{\xi_{\bf p}e^{iprx}}{E^2 - \xi_{\bf p}^2 - \Delta^2 } \frac{\omega_D^2}{\xi_{\bf p}^2 +     \omega_D^2}.
\end{eqnarray}
Note that we introduced a convergence factor $\omega_D^2/(\xi_{\bf p}^2+\omega_D^2)$ into $I_1$. While the integral $I_0$ is automatically dominated by the vicinity of the Fermi surface, this is not the case for $I_1$ in the absence of the convergence factor. In that case, we need to account for the fact that the BCS model underlying the calculations is restricted to energies smaller than the Debye frequency $\omega_D$. We will ultimately restrict attention to $r\gg v_F/\omega_D$. In this limit, we can formally eliminate the cutoff from the result by taking the limit $E_F,\omega_D \to \infty$ while keeping $\omega_D/E_F\ll 1$. Note that an accurate theory for $r\ll v_F/\omega_D$ would require one to develop a more microscopic theory of the underlying superconductor. 

To evaluate the integrals, we first perform the integral over $\xi_{\bf p}$ and subsequently the $x$-integration. Writing $p = p_F + \xi_{\bf p}/v_F$, this can be done straightforwardly for $I_0$ and we obtain
\begin{equation}
  I_0 = -\frac{\pi \nu_0}{\sqrt{\Delta^2 - E^2}} \frac{\sin k_F r}{k_F r} e^{-\sqrt{\Delta^2 - E^2}r/v_F}.
\end{equation} 
Similarly, we can evaluate the integral for $I_1$. Taking the limit described above, we find 
\begin{equation}
  I_1 = -\pi \nu_0 \frac{\cos k_F r}{k_F r} e^{-\sqrt{\Delta^2 - E^2}r/v_F}.
\end{equation} 
More explicitly, the corrections to this result either decay exponentially with $r$ on the scale $v_F/\omega_D$ or are suppressed as powers of $\omega_D/E_F$, and thus vanish after taking the limit. 

\section{Momentum-space Bogoliubov-de Gennes Hamiltonian}
\label{hdelta}

In this Appendix, we sketch the derivation of Eqs.\ (\ref{hk}) and (\ref{deltak}). Inserting Eq.\ (\ref{heff}) into Eq.\ (\ref{hfourier}), we readily find Eq.\ (\ref{hk}) with 
\begin{equation}
  F(k) = - \Delta {\rm Im} \sum_{j=1}^\infty \frac{1}{k_F a j} e^{-a j/\xi_0} [e^{i(k_F+k)aj}+e^{i(k_F - k)aj}].
\end{equation}
Here, we have dropped the trivial term involving the Shiba energy $\epsilon_0$ of the individual impurities. The sum over $j$ can be readily performed by the identity
\begin{equation}
   - \ln (1-x) = \sum_{j=1} \frac{x^j}{j},
   \label{ln}
\end{equation}
which yields
\begin{eqnarray}
  F(k) &=&  \frac{\Delta}{k_F a} {\rm Im}\left [ \ln ( 1-  e^{-a/\xi_0 + i (k_F+k)a}) \right.
  \nonumber\\
 && \left. + \ln ( 1-  e^{-a/\xi_0 + i (k_F - k)a}) \right] .
\end{eqnarray}
Finally, we use the identity ${\rm Im}\ln z = i\arctan ({\rm Im}z/{\rm Re}z)$ to obtain the result given above in Eq.\ (\ref{Fk}).

Similarly, inserting Eq.\ (\ref{deltaeff}) into Eq.\ (\ref{deltafourier}), we find Eq.\ (\ref{deltak}) for $\Delta_k$ with 
\begin{equation}
  f(k) =  \sum_{j=1}^\infty \frac{e^{-ja/\xi_0}}{j}[e^{ikaj} + e^{-ikaj}].  
\end{equation}
Performing the sum over $j$ using Eq.\ (\ref{ln}) yields Eq.\ (\ref{fk}) of the main text.


\begin{thebibliography}{12}
\bibitem{footnote} The chirality of the helix is probably determined by the Dzyaloshinskii-Moriya component of the interaction. This component breaks the degeneracy between the two helicities and may be much smaller than the isotropic part $J(i-j)$ of the RKKY interaction.
\bibitem{foot1} However, our conclusions remain qualitatively valid even without this assumption as illustrated by the numerical results in Fig.\ \ref{fig:small_xi}.
\bibitem{footnote1} One can only write the wavefunction $\psi_1$ in terms of real Majorana wavefunctions, which are located at only one end, in the case of $\theta=\pi/2$. For arbitrary $\theta$ the phase of $\psi_1$ varies spatially due to the complex long-range hopping terms and one cannot transform to a Majorana basis by a global rotation. We checked numerically that the decay of the wavefunction also obeys a $1/[x\ln^2(x/x_0)]$ law for $\theta<\pi/2$.
\end{thebibliography}
\end{document}